\documentclass[a4paper,12pt,nofootinbib,superscriptaddress,showpacs]{revtex4}
\usepackage[frenchb,english]{babel}
\usepackage[applemac]{inputenc}
\usepackage{enumerate}
\usepackage{amsmath}
\usepackage{amsthm}
\usepackage{amssymb}
\usepackage{graphicx}
\usepackage{fancybox}
\usepackage{color}
\usepackage{hyperref}

\renewcommand{\Re}{\textrm{Re}}

\newcommand{\be}{\begin{equation}}
\newcommand{\ee}{\end{equation}}
\newcommand{\bal}{\begin{equation} \begin{aligned}}
\newcommand{\eal}{\end{aligned} \end{equation}}
\newcommand{\bi}{\begin{itemize}}
\newcommand{\ei}{\end{itemize}}
\newcommand{\ben}{\begin{enumerate}}
\newcommand{\een}{\end{enumerate}}
\newcommand{\bmat}{\begin{pmatrix}}
\newcommand{\emat}{\end{pmatrix}}

\newcommand{\Om}{\Omega}
\newcommand{\om}{\omega}

\newcommand{\eq}[1]{Eq.~\eqref{#1}}

\begin{document}
\baselineskip 17pt

\selectlanguage{english}

\title{Hawking radiation of massive modes and undulations}

\author{Antonin~Coutant}
\email{antonin.coutant@th.u-psud.fr}
\affiliation{Laboratoire de Physique Th\'eorique, CNRS UMR 8627, B\^at. 210, Universit\'e Paris-Sud 11, 91405 Orsay Cedex, France}
\author{Alessandro~Fabbri}
\email{afabbri@ific.uv.es}
\affiliation{Departamento de F\'isica Te\'orica and IFIC, Universidad de Valencia-CSIC, C. Dr. Moliner 50, 46100 Burjassot, Spain}
\author{Renaud~Parentani} 
\email{renaud.parentani@th.u-psud.fr}
\affiliation{Laboratoire de Physique Th\'eorique, CNRS UMR 8627, B\^at. 210, Universit\'e Paris-Sud 11, 91405 Orsay Cedex, France}
\author{Roberto~Balbinot}
\email{Roberto.Balbinot@bo.infn.it}
\affiliation{Dipartimento di Fisica dell’Universit\'a di Bologna and INFN sezione di Bologna, Via Irnerio 46, 40126 Bologna, Italy}
\author{Paul~R.~Anderson}
\email{anderson@wfu.edu}
\affiliation{Department of Physics, Wake Forest University, Winston-Salem, North Carolina 27109, USA}

\date{\today}

\begin{abstract}
We compute the analogue Hawking radiation for modes which possess a small wave vector perpendicular to the horizon.
For low frequencies, the resulting mass term induces a total reflection. 
This reflection is accompanied by an extra mode mixing which occurs in the
supersonic region, and which cancels out the infrared divergence of the near horizon spectrum.
As a result, the amplitude of the undulation (0-frequency wave with macroscopic amplitude) 
emitted in white hole flows now saturates at the linear level, unlike what is found in the massless case.
In addition, we point out that the mass introduces a new type of undulation
which is produced in black hole flows, and which is well described in the
hydrodynamical regime.
\end{abstract}

%\keywords{Hawking radiation, massive fields, undulation, Bose-Einstein condensate}
%revtex4 - option : showkeys

\hfill \small LPT-Orsay 12-59

\maketitle

\newpage

\section*{Introduction}

Recent studies of the analogue Hawking radiation~\cite{Unruh81} have shown
that a standing (zero-frequency) wave is emitted in the
supersonic region of white hole flows~\cite{Mayoral2011,ACRPSF}.
This wave possesses a macroscopic amplitude and a short wavelength
fixed by the dispersive properties of the medium.
Interestingly, it corresponds to some well known solutions in hydrodynamics~\cite{Hydro}~\footnote{
Even though these are well known, they are not completely understood.
In the Introduction of Ref.~\cite{Thesis},
one reads:  `Still, the characteristics and the formation of an undular hydraulic jump are
not fully understood'.}, namely undulations associated with hydraulic jumps.
In addition, undulations have been recently observed in water tank experiments~\cite{Rousseaux,SilkePRL2010}
aiming to detect the analogue Hawking radiation,
but their relation with the Hawking effect was
not pointed out. This relation was understood in 
the context of atomic Bose-Einstein condensates (BEC), where
the emission of an undulation was explained in terms of a combination of several effects.
Firstly, $n_\om$, the spectrum of massless phonons spontaneously produced {\it \`a la Hawking}
from the sonic horizon diverges like $1/\om$ for $\om \to 0$,
as in the Planck distribution. Secondly, $p_U = p_{\om = 0}$,
the wave number of the undulation is a non-trivial solution of the dispersion relation, and thirdly,
its group velocity is oriented away from the horizon.

\begin{figure}[!h]
\begin{center}
\includegraphics[scale=0.65]{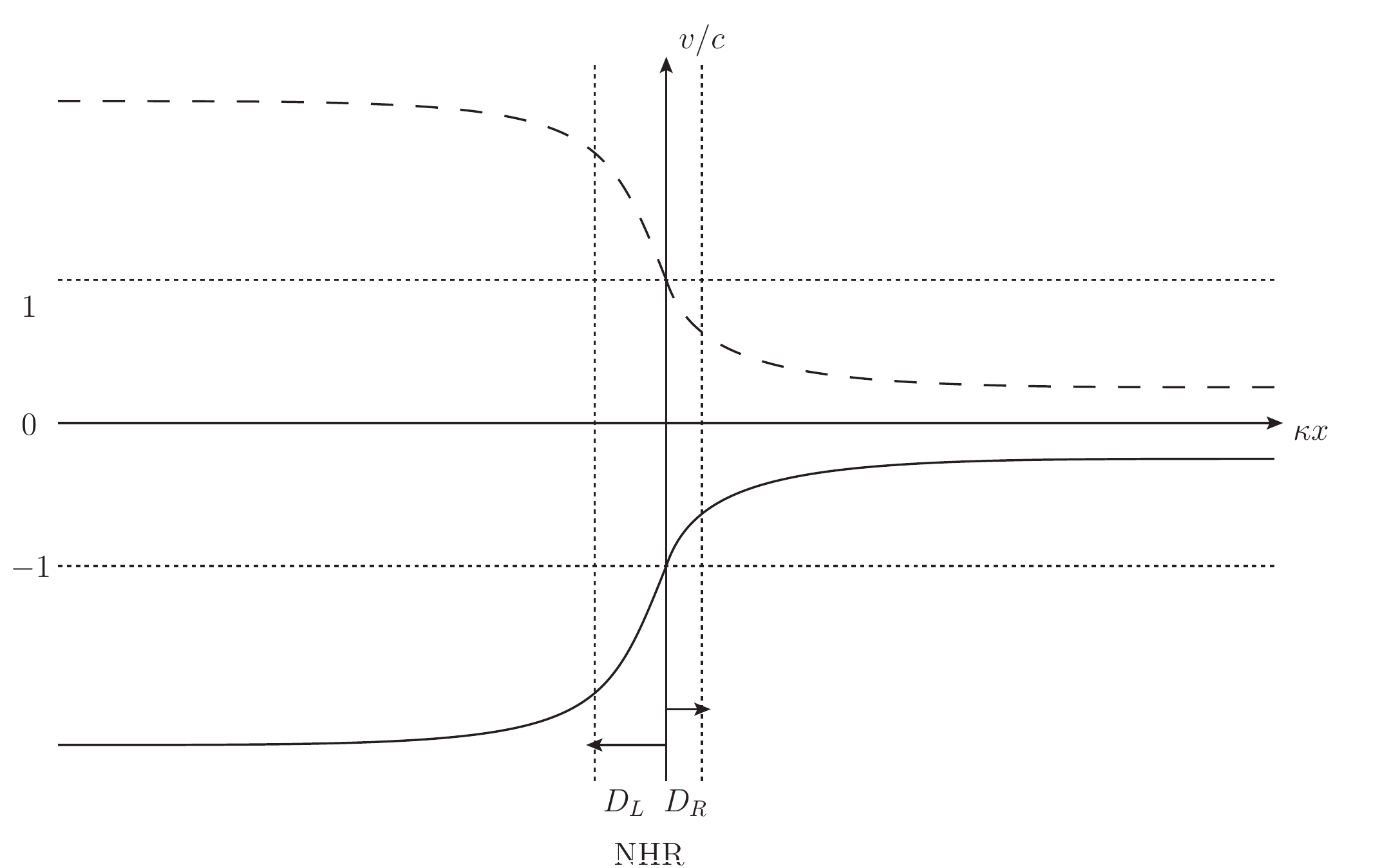}
\end{center}
\caption{Shown in this figure are examples of one dimensional black hole flow (solid line) and white hole flow (dashed line)
with regular asymptotic properties. They are related to each other
by reversing the sign of the velocity $v(x) \to -v(x)$.
In both cases, the subsonic R region $\vert v \vert < c = 1$
is on the right of the horizon, while the supersonic L region is on the left.
The near horizon region (NHR), where $v \sim -1 + \kappa x$  is a good approximation,
has a width in units of $\kappa$ of $D_L$ on the left and of $D_R$ on the right. }
\label{vprofile_fig}
\end{figure}

\begin{figure}[!h]
\begin{center}
\includegraphics[scale=0.8]{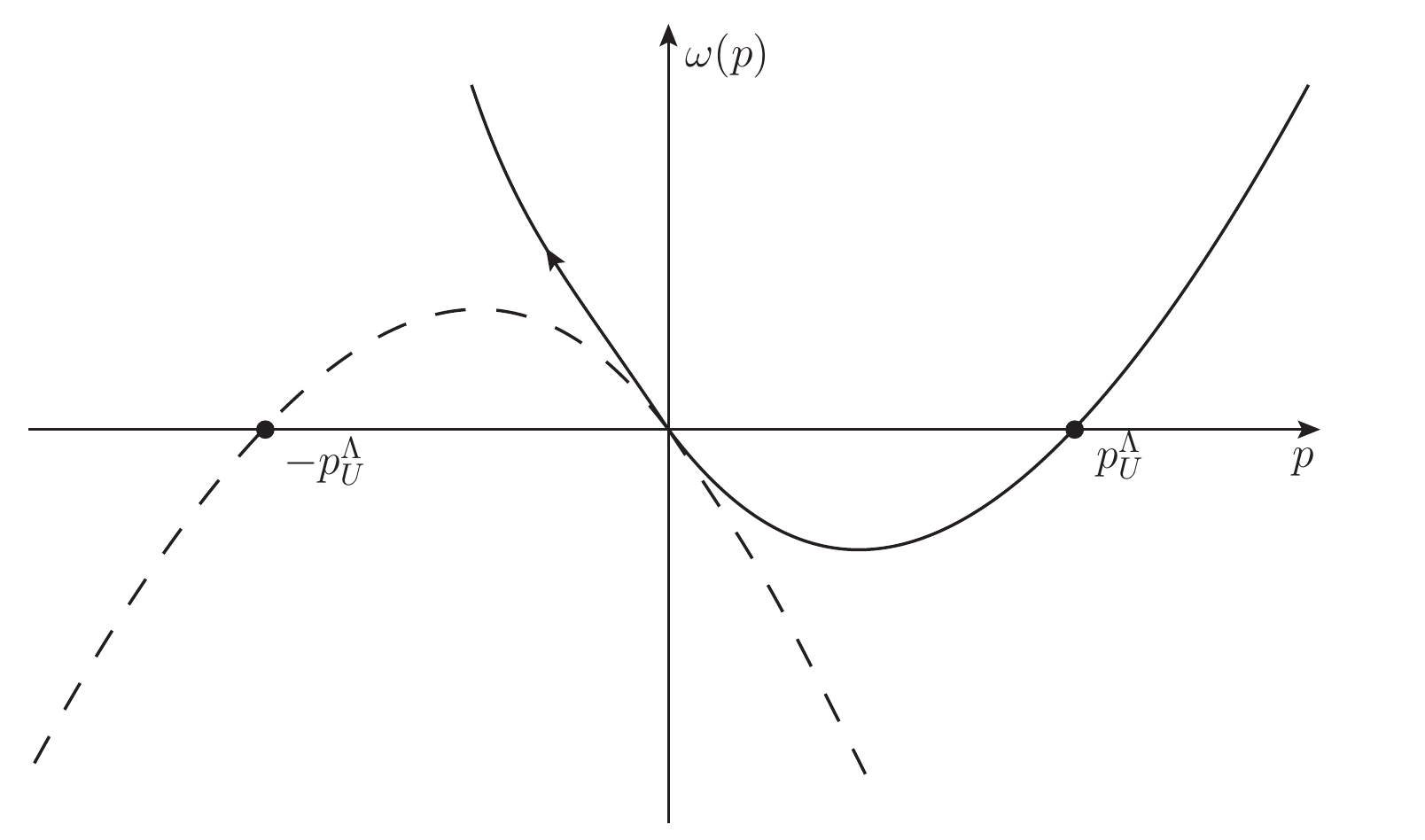}
\end{center}
\caption{The solid line represents the positive branch of the comoving frequency $\Om$, while the
dashed line represents the negative branch. One clearly sees that the superluminal Bogoliubov dispersion
is responsible for the two zero-frequency roots $\pm p^\Lambda_U$. The sign of the group velocity
can be seen from the slope of the solid line at the corresponding root.}
\label{WHdisprel_fig}
\end{figure}

To see this in more detail first note that the Bogoliubov dispersion relation
in a one dimensional stationary flow, and for a longitudinal wave vector $p$, is
\be
\Om = \om - v p = \pm \sqrt{c^2 p^2 (1 + \xi^2 p^2)} ,
\label{Bogdr}
\ee
where $\om$ is the conserved frequency, $v$ is the flow velocity, $c$ is the speed of sound, and $\xi = \hbar/2m_{\rm at} c$ is the
healing length, given in terms of the mass of the atoms $m_{\rm at}$.  
The $\pm$ sign refers to positive and negative norm branches, see {\it e.g.}~\cite{Mayoral2011} for details. The flow profiles giving rise to a black hole (BH)
and a white hole (WH) sonic horizon are represented in Fig.~\ref{vprofile_fig}.
The dispersion relation of \eq{Bogdr} evaluated in the asymptotic
supersonic region is plotted in Fig.~\ref{WHdisprel_fig}.
 The zero-frequency roots are $\pm p^\Lambda_U$, where $p^\Lambda_U$ is
\be
p^\Lambda_U = \Lambda\,  \sqrt{v_L^2 - c_L^2} .
\label{pU}
\ee
In this equation,
 $v_L$ and $c_L$ are the asymptotic values of the velocity and speed of sound in the
supersonic region L, and $\Lambda = 1/c_L \xi_L =2m_{\rm at}/\hbar$ characterizes the short distance dispersion.
This root only exists in a supersonic flow, and its associated group velocity
$v_{\rm gr} = 1/\partial_\om p$
is directed {\rm against} the flow. Hence in a BH flow, it is oriented towards the horizon, whereas
for a WH one it is oriented away from it.
This explains why the zero-frequency mode only appears in WH flows
where it is generated  at the sonic horizon.
At this point it should be mentioned that these solutions are not restricted to superluminal dispersion.
A completely similar phenomenon exists in fluids characterized by a subluminal dispersion relation, such as that obtained by replacing $\xi^2 \to -\xi^2$ in  \eq{Bogdr}.
This time however, the zero frequency root, and the corresponding undulation,
live in the subsonic R region of the WH flow. This can be understood because
of the (approximate) symmetry of the mode equation found in~\cite{ACRPSF}, 
which replaces a superluminal dispersion by a subluminal one.

\begin{figure}[!h]
\begin{center}
\includegraphics[scale=0.8]{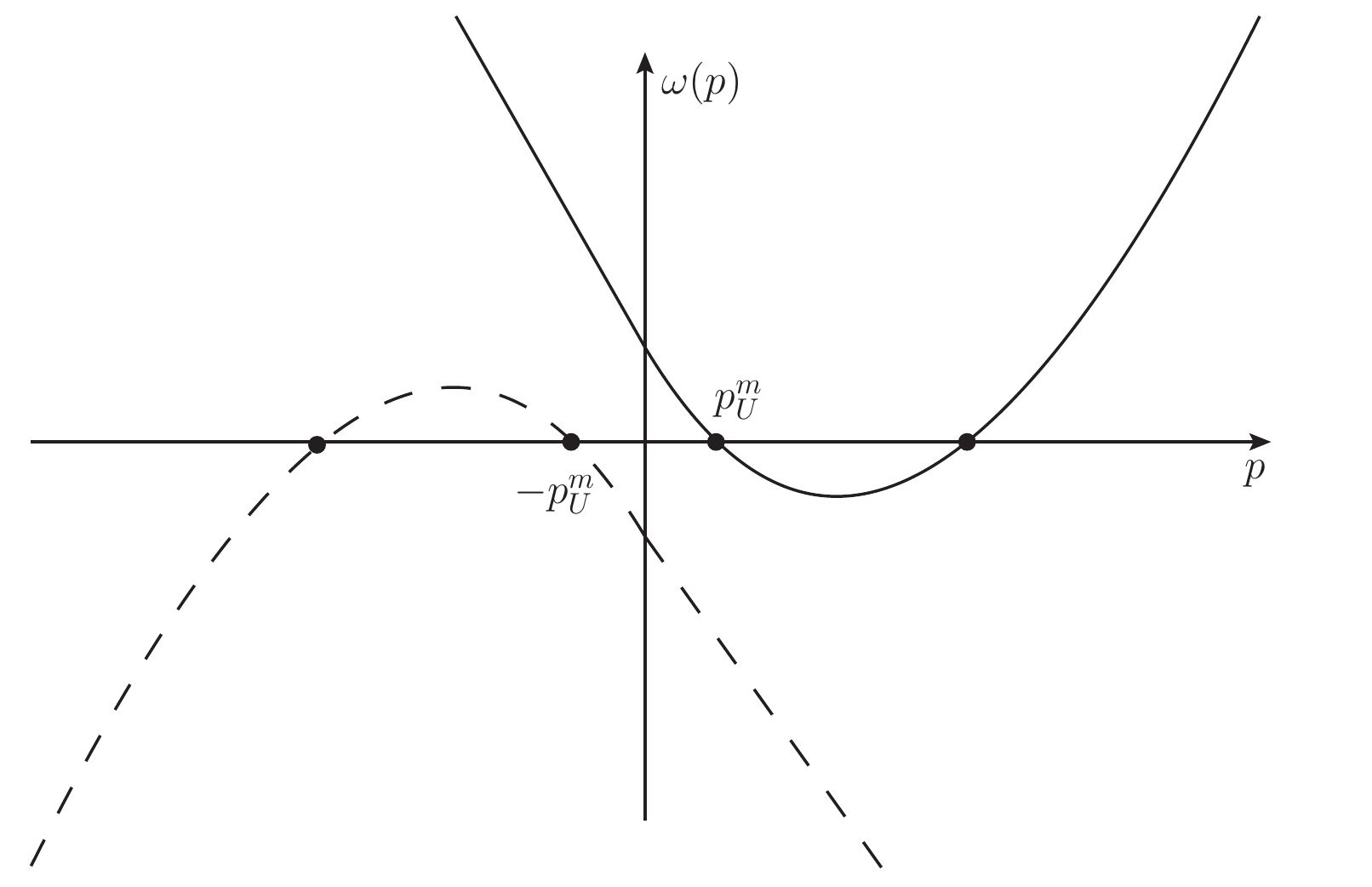}
\end{center}
\caption{As in Fig.~\ref{WHdisprel_fig}, the solid line represents the positive branch of the comoving frequency $\Om$, while the
dashed line represents the negative branch.
One sees that $p_\perp^2$, which acts as a mass, is responsible for new zero-frequency roots $\pm p^m_U$ which occur in the phonon part of
the dispersion relation. }
\label{WHdisprelm_fig}
\end{figure}

When considering elongated quasi one dimensional systems,
but relaxing the assumption that the phonon excitations are purely longitudinal, the
phonon modes are now characterized by  their transverse wave number $p_\perp$, which takes discrete values $2 \pi n/L_\perp$, where $n$
is an integer and $L_\perp$ is the characteristic size of the perpendicular dimensions. When
$p_\perp^2 \neq 0$, the modified dispersion relation replacing \eq{Bogdr} is
\be
\Om = \om - v p = \pm \sqrt{c^2 (p^2 +p_\perp^2)(1 + \xi^2 (p^2 +p_\perp^2))} \, .
\label{BogdrNl}
\ee
It is represented in Fig.~\ref{WHdisprelm_fig}.
When $p_\perp^2 \xi^2 \ll 1$, $c^2 p_\perp^2$ acts as a mass squared.
In this regime, there are two new zero-frequency
roots $\pm p^m_U$. They live in the hydrodynamical regime, characterized by a relativistic
linear dispersion relation.
Indeed, in the limit $\xi^2 p_\perp^2 \to 0$, and
if $v_L^2/c_L^2$ is not too close to 1, $ p^m_U$ is independent of $\xi$ and given by
\be
p^m_U = \frac{c_L p_\perp}{\sqrt{v_L^2- c_L^2 }} .
\label{pUm}
\ee
In addition we note that the group velocity of this new solution has a sign
opposite to that of \eq{pU}. Hence, this new solution
will be emitted in BH flows but not in WH ones.

In brief, we see that the introduction of a perpendicular momentum opens the
possibility of finding `massive' undulations in BH flows, which are well-described
in the hydrodynamical approximation of the underlying condensed matter system.
To verify if this is the case, one should see how the mass affects the spectrum, and in particular
if it acts as an infrared regulator that saturates the growth of the undulation amplitude found in the massless case~\cite{Mayoral2011,ACRPSF}.
In this paper, these issues will be investigated in a simplified context where the phonon modes
obey a second order differential equation which is a massive Klein-Gordon equation in a curved metric.
This analogy should work not only for BEC but for other condensed matters systems
where the quasi-particle dispersion relation is linear at low frequency. 
We notice that similar issues related to the Cerenkov effect have been recently discussed in~\cite{Caru_Rouss}.

This paper is organized as follows. In Sec.~\ref{Sett}, we study the solutions of
the Klein-Gordon equation in a stationary BH metric. We explain how the $in/out$
scattering matrix can be decomposed into three blocks that each encode some
aspect of mode mixing of massive fields.
In Sec.~\ref{quattro}, we study three preparatory cases
which are then combined so as to obtain the $S$-matrix in a black hole flow
similar to that represented in Fig.~\ref{vprofile_fig}.
In Sec.~\ref{UndulSec}, we study the properties of the two-point correlation function
in the low frequency sector, and its relationship with the undulations.

\section{Settings, mode mixing, structure of the $S$-matrix}
\label{Sett}

In this paper, we study the behavior of a massive scalar field in the 1+1 dimensional metric
\be
ds^2 = c^2 dt^2 - (dx - v(x) dt)^2 . \label{PGds}
\ee
For simplicity we impose the condition that the speed of sound is constant,
and work in units where $c^2 = 1$.
The stationary metric of \eq{PGds} possesses a Killing field $K_t = \partial_t$,
whose norm is $1-v^2$. It is time-like in subsonic flows,
space-like in supersonic ones, and it vanishes when $1-v^2$ crosses 0. At that place, there is a horizon. In what follows, we work with $v < 0$ and with
monotonic flows $1 - v^2$ that possess a horizon at $x=0$, where $v=-1$.
The definition of the surface gravity of the horizon we shall use is
\be
\kappa = \frac12 \partial_x (1 - v^2)_{|0}. \label{kappadef}
\ee
We adopt this local definition, which no longer refers to the norm of $K_t$ at infinity, because it
allows us to compare various geometries starting from the near horizon region (NHR).
For a black hole, we have $\kappa >0$, while for a white hole $\kappa < 0$.
Unless specified otherwise, we shall only consider black holes.
Notice also that we shall work with flow velocities that are either asymptotically bounded or unbounded; in the latter case, there will be a singularity.

The field will be studied at fixed Killing frequency $\om = - (K_t)^{\mu} P_\mu$, using a decomposition into stationary modes
\be
\phi = \int \phi_\om(x) e^{-i \om t} d\om.
\ee
At fixed $\om$, the Klein-Gordon
equation gives
\be
[ (\om + i\partial_x v)(\om + iv\partial_x) + \partial_x^2 - m^2] \phi_\om(x) = 0 \label{modequ}.
\ee
Similar equations are obtained when studying acoustic perturbations on a fluid flow with a velocity profile $v(x)$~\cite{Unruh95,Rivista05}.
In these cases, a non zero transverse momentum $p_{\perp}$ plays the role of
the mass $m$. In App.~\ref{metricsApp}, we present the various (effectively) massive wave equations
and their differences. In what follows, we consider only
\eq{modequ}, for profiles $v(x)$ that give rise to analytically soluble equations.
Yet, we aim to extract generic features. When studying numerically the phonon mode equation in
a Bose condensate and with a varying sound speed~\cite{to-appear},
we recovered the features found for solutions to \eq{modequ}.

\subsection{Classical trajectories}
\label{clatrajSec}

To understand the consequences of the mass on black hole radiation, it is useful to first consider
the corresponding classical problem where $p = (\partial_x)^\mu P_\mu$ is the momentum of the massive particle at fixed $\om$.
In that case, the Hamilton-Jacobi equation associated with \eq{modequ} is
\be
\Om^2 = (\om - v(x) p)^2 = p^2 + m^2 \label{HJ},
\ee
where $\Om = \om - v p$ is the comoving frequency. \eq{HJ} admits two roots
\begin{align}
p_\pm &= \frac{-\om v \pm \sqrt{\om^2 - m^2(1-v^2)}}{1-v^2}, \label{roots}
\end{align}
The classical trajectories obey Hamilton's equations $dx/dt = 1/\partial_\om p$
and $dp/dt = - 1/\partial_\om x$. We summarize here their main features with $\om > 0$,
see \cite{TJRPentrop,Boom,Jannes11} for more details.
\bi
\item Close to the horizon, at first order in $1-v^2 \sim 2\kappa x \ll 1$, one has
\begin{align}
p_+ &= \frac{\om}{\kappa x},\\
p_- &= \frac{m^2 - \om^2}{2\om} .
\end{align}
We see that $p_+$ diverges for $x\to 0$ whereas $p_-$ hardly varies.
The corresponding geodesics follow
\begin{align}
x_+ (t) &= x_+^0 \, e^{\kappa t},\\
x_- (t) &= x_-^0 - \frac{2\om^2}{\om^2 + m^2} t. 
\end{align}
The second trajectory is regularly falling
across the horizon, while the first undergoes an infinite focusing in the past, in a mass independent manner.

\item Far away from the horizon,  in the left region (L) for $1-v^2 < 0$,
both solutions are moving to the left (since $v< 0$) even though $p_+<0$ and $p_- >0$.

\begin{figure}[!h]
\begin{center}
\includegraphics[scale=1]{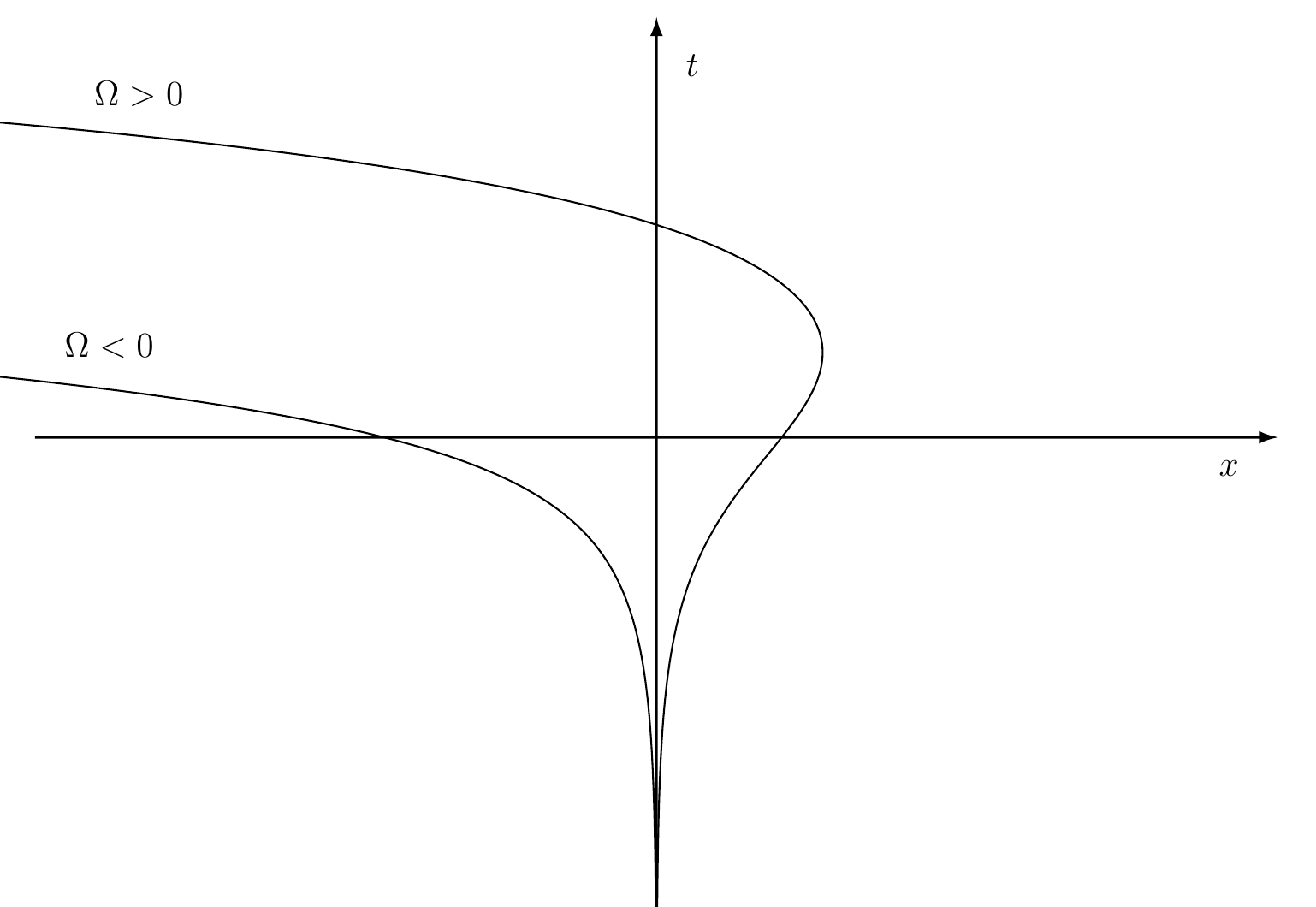}
\end{center}
\caption{In this figure trajectories of massive particles for a fixed frequency $\om > 0$ below the threshold of \eq{thres} are shown.
The positive momentum trajectory is reflected in the outside region at the turning point $x_{\rm tp}(\om)$
whereas the negative momentum one (or equivalently the positive momentum one with negative $\om$~\cite{TJ96})
propagates in the inside region
where the Killing field is space-like. The negative momentum particle
has a negative comoving frequency $\Om = \om - vp$, and corresponds to a negative norm mode,
as we shall see later in the text.}
\label{mass_traj_fig}
\end{figure}

\item For $1-v^2 >0$, in the right region (R), as long as $(1-v^2)m^2 < \om^2$, there are two real roots.
At some point $x_{\rm tp}$ we reach $(1-v^2)m^2 = \om^2$ where they
become complex. This means that the trajectory is
reflected and falls back across the horizon.
Hence, the asymptotic value of $1 - v^2 > 0$ determines
the threshold frequency
\be
\om_R = m\sqrt{1 - v_{\rm as}^2}, \label{thres}
\ee
above which the trajectory is not reflected.
When $\om < \om_R$, there is a single trajectory with $p_\om > 0$
that starts from the horizon to the right and bounces back across the horizon,
see Fig.\ref{mass_traj_fig}. For $\om > \om_R$ instead, there are 2 disconnected
trajectories, one is moving outwards from the horizon, while the other falls in from $x = \infty$.
As we shall see,
the dimensionality of asymptotic modes will be different above and below $\om_R$.
\ei

\subsection{Mode mixing}

We decompose the field operator in a basis of stationary modes
\be
\hat \phi(t,x) =\sum_j \int_0^{+\infty} \left[ \hat a_\om^j \phi_\om^j(x) + \hat a_\om^{j \dagger} (\phi_{-\om}^j(x))^* \right] e^{-i\om t} d\om + h.c.\, , \label{phidecomp}
\ee
where the discrete index $j$ takes into account the dimensionality of mode basis at fixed $\om$.
The basis is orthonormal in the sense of the Klein-Gordon scalar product
\begin{align}
(\phi_\om^j|\phi_{\om'}^{j'}) &= \int_{\mathbb R} \left[ \phi_\om^{j*}(\om' + i v\partial_x)\phi_{\om'}^{j'} + \phi_{\om'}^{j'} (\om - i v\partial_x) \phi_\om^{j*}\right] dx \label{scalprod}, \nonumber
\\
&= \pm \delta(\om - \om') \delta_{jj'}.
\end{align}
Following the standard conventions~\cite{Wald}, we name the negative norm modes $(\phi_{-\om})^*$
so that $e^{i \om t}\phi_{-\om}$ is a positive norm mode with negative frequency.

To obtain the dimensionality of the mode basis, one must identify
the solutions of \eq{modequ} that are asymptotically bounded modes (ABM).
This requirement univocally picks out a \emph{complete} basis over which the canonical field $\hat \phi$ must be decomposed~\cite{MacherWH}.
Asymptotically, solving the mode equation \eqref{modequ} is equivalent to solving the Hamilton-Jacobi equation \eqref{HJ}.
Hence, the dimensionality of the ABM can be found
by considering the real roots of \eq{HJ}.
Moreover, the sign of the norm of an asymptotic mode
is given by the sign of the corresponding comoving
frequency $\Om(p_i)= \om - v p_i$, as can be seen from \eq{scalprod}.

In addition, because the situation is non-homogeneous, modes mix and the basis is not unique.
As usual, we introduce {\it in} modes $\phi_\om^{\rm in}$ and {\it out} modes $\phi_\om^{\rm out}$ by examining the mode behavior at early and late times;
see the discussion after \eq{regularcondition} for more precision. The $S$-matrix then relates the {\it in} and {\it out} bases. When there is a
horizon, these two basis are inequivalent because positive and negative norm modes coexist and mix.
To further study the mixing, one should consider separately the frequencies below and above $\om_R$
in \eq{thres}.

For $0 < \om < \om_R$, there are two ABM.
One has a negative norm and propagates behind the horizon.
 The other has a positive norm, comes out from the horizon, and bounces back
across the horizon, see Fig.~\ref{mass_traj_fig}. The $S$-matrix thus has the form
\be
\bmat \phi_\om^{\rm in} \\ \left(\phi_{-\om}^{\rm in}\right)^* \emat = S^T \cdot \bmat \phi_\om^{\rm out} \\ \left(\phi_{-\om}^{\rm out}\right)^* \emat .
\label{Sdef22}
\ee
To follow the standard definition of the $S$-matrix,
we use its transpose here.

For $\om > \om_R$, there are three ABM. The negative norm one still propagates behind the horizon.
The second one has a positive norm, comes out from the horizon, and reaches infinity. The third one comes from infinity and falls into the hole.
As in~\cite{MacherWH}, we denote the first two with the superscript $u$ and the last one with $v$,
because at high momentum, when the mass is negligible, they follow
retarded ($u$) and advanced ($v$) null geodesics.
We then define $S$ by
\be
\bmat \phi_\om^{\rm in, u} \\ \left(\phi_{-\om}^{\rm in, u}\right)^* \\ \phi_\om^{\rm in, v} \emat = S^T \cdot \bmat \phi_\om^{\rm out, u} \\ \left(\phi_{-\om}^{\rm out, u}\right)^* \\ \phi_\om^{\rm out, v} \emat.
 \label{Sdef33}
\ee
In this regime,  when starting from vacuum,
the three {\it out} occupation numbers $n^{\rm u}_\om$, $n^{\rm v}_\om$
and $n^{\rm u}_{-\om}$ obey $n^{\rm u}_\om + n^{\rm v}_\om = n^{\rm u}_{-\om}$
because of the stationarity of the settings. The first two are given by the square of the overlaps
\be
n^{\rm u}_\om = |(\phi_\om^{\rm out, u} | \phi_{-\om}^{\rm in, u*})|^2, \qquad
n^{\rm v}_\om = |(\phi_\om^{\rm out, v} | \phi_{-\om}^{\rm in, u*})|^2. \label{occupnum+}
\ee
As we shall see, in both cases, it is useful to decompose the total $S$-matrix as
 \be
S = S_{\rm far} \cdot S_{\rm ext} \cdot S_{\rm NHR}, \label{Sfactogene}
\ee
where each $S$-matrix describes one step of the {\it in/out} scattering.
The first one, $S_{\rm NHR}$, describes the mode mixing which arises for high momenta $p \gg m$,
near the horizon where the modes are effectively massless. The second matrix $S_{\rm ext}$ encodes the elastic scattering which occurs in the external region R.
Below the threshold, it describes the total reflection, while above it governs the
grey body factors encoding the partial transmission.
The last matrix $S_{\rm far}$ describes the mixing occurring in the left region
between the two modes that are propagating towards $x = - \infty$.

It should be mentioned that this decomposition is not unique, as only the {\it in/out} $S$-matrix
is univocally defined. However, in the absence of dispersion, each $S$-matrix is solution of a well defined and independent scattering problem.  
In addition, this decomposition is very useful as it allows us to compute $S$, and to understand its properties. 

\subsection{Near horizon scattering}
\label{NHRSec}

We start with $S_{\rm NHR}$ because its properties are valid for all metrics possessing
a horizon and because they are determined for momenta much higher than the mass
and in the immediate vicinity of the horizon.
To simplify the mode equation (\ref{modequ}), we introduce the
auxiliary mode $\varphi_\om$
\be
\phi_\om(x) = \frac{e^{-i \om \int^x \frac{v(x')}{1-v^2(x')}dx'}}{\sqrt{|1-v^2|}} \varphi_\om(x). \label{def_varphi}
\ee
\eq{modequ} is then cast in a canonical form, without the term linear in $\partial_x$,
\be
\left[ - \partial_x^2 + \left(\frac{\partial_x^2\sqrt{|1-v^2|}}{\sqrt{|1-v^2|}} +\frac{m^2}{1-v^2} -\frac{\om^2}{(1-v^2)^2}\right) \right] \varphi_\om(x) = 0 . \label{canmodequ}
\ee
We notice that the norm of $\varphi_\om$ is, up to a sign,
given by the Wronksian
\be
W(\varphi) = 2i\pi \left(\varphi_\om^* \partial_x \varphi_\om - \varphi_\om \partial_x\varphi_\om^* \right) \label{Wronskian}.
\ee
Using \eq{def_varphi}, one verifies that unit Wronskian $\varphi_\om$ modes,
give rise to $\phi_\om$ modes that have a unit norm with respect to the scalar product of \eq{scalprod}.
The relative sign is given by that of the comoving frequency $\Om$ of \eq{HJ}.

In the close vicinity of the horizon,  the mass term becomes negligible in \eq{canmodequ}.
More precisely, in the near horizon region where $1-v^2 \sim 2\kappa x$,
keeping only the leading term for $\kappa x \ll 1$, one obtains
\be
\left[ - \partial_x^2 -\left( \frac14 + \frac{\om^2}{4\kappa^2}\right)\frac1{x^2} \right] \varphi_\om(x) = 0 .
\ee
Therefore the leading behavior of $\varphi_\om$ is
\begin{align}
\varphi_\om \underset{x \to 0}{\sim} \, &\underbrace{\Theta(-x)\, A |2\kappa x|^{i \frac{\om}{2\kappa} +\frac12}}_{\text{focusing on the left}} + \underbrace{\Theta(x)\ A' \, |2\kappa x|^{i \frac{\om}{2\kappa} +\frac12}}_{\text{focusing on the right}} \nonumber \\
&+ \underbrace{\Theta(-x)\, B  |2\kappa x|^{-i \frac{\om}{2\kappa}+\frac12}  +\Theta(x) \, B'  |2\kappa x|^{-i \frac{\om}{2\kappa} +\frac12}}_{\text{regularly falling in}}. \label{NHRbehav}
\end{align}
When re-expressing this in terms of the original mode $\phi_\om$ using \eq{def_varphi},
we see that the $B$ weighted terms are regular, and are in fact constant.
Therefore, they account for the regularity of the left moving mode as it crosses the horizon. Hence, we impose
\be
B=B'. \label{regularcondition}
\ee
On the other hand, the $A$ parts in \eq{NHRbehav} oscillate infinitely around $x=0$, and account for
high momentum modes living on either side of the horizon, and which are singular on it.
As understood by Unruh~\cite{Unruh76}, it is appropriate to combine them
in superpositions that are analytic either in the upper, or lower, half complex $x$-plane.
The reason is that their analytical properties guarantee that these modes correctly
characterize the stationary vacuum state which is regular across the horizon, and which plays the role of the $in$ vacuum.
This characterization applies to the $\phi_\om$ modes which are solutions to \eq{modequ}.
Hence, the modes $\varphi_\om$ of \eq{def_varphi} are products of a non analytic function and an analytic one, which is an Unruh mode:
\begin{align}
\varphi_\om^{\rm in}
&\sim \Gamma \left(i\frac{\om}{\kappa}\right) \frac{e^{\frac{\om \pi}{2\kappa}}}{\sqrt{8\pi^2 \kappa}} |x|^{-i\frac{\om}{2\kappa}+\frac12}\times \left( x+i\epsilon \right)^{i\frac{\om}{\kappa}}, \label{inplus}\\
\left(\varphi_{-\om}^{\rm in}\right)^*
&\sim \Gamma \left(i\frac{\om}{\kappa}\right) \frac{e^{\frac{\om \pi}{2\kappa}}}{\sqrt{8\pi^2 \kappa}} |x|^{-i\frac{\om}{2\kappa}+\frac12} \times \left( x-i\epsilon \right)^{i\frac{\om}{\kappa}}. \label{inminus}
\end{align}
We have used \eq{Wronskian} to normalize these modes, and their phases have been chosen in order to obtain simple expressions.
When there are turning points, as is the case for dispersive fields~\cite{ACRPSF} and for massive fields, one should pay attention to these phases.

The normalized modes that propagate on either side of the horizon and vanish on the other side are
\begin{align}
\varphi_\om^{\rm Right} &\sim \Theta(x) \frac{|2\kappa x|^{i\frac{\om}{2\kappa} +\frac12}}{\sqrt{4\pi \om}},\label{rightm}\\
\left(\varphi_{-\om}^{\rm Left}\right)^* &\sim \Theta(-x) \frac{|2\kappa x|^{i\frac{\om}{2\kappa} +\frac12}}{\sqrt{4\pi \om}}.
\end{align}
The near horizon $S$-matrix $S_{\rm NHR}$  is then defined by
\be
\bmat \phi_\om^{\rm in} \\ \left(\phi_{-\om}^{\rm in}\right)^* \emat = \bmat \alpha_\om^{\rm NHR} & \tilde \beta_\om^{\rm NHR} \\ \beta_\om^{\rm NHR} & \tilde \alpha_\om^{\rm NHR} \emat \cdot \bmat \phi_\om^{\rm Right} \\ \left(\phi_{-\om}^{\rm Left}\right)^* \emat.
\label{SNHR_1}
\ee
Using the analytic properties of the {\it in} modes of Eqs. \eqref{inplus}, \eqref{inminus}, we immediately obtain
\be
S_{\rm NHR} = \sqrt{\frac{\om}{2\pi \kappa}} \Gamma \left(i\frac{\om}{\kappa}\right) \bmat e^{\frac{\om \pi}{2\kappa}} & e^{-\frac{\om \pi}{2\kappa}} \\ e^{-\frac{\om \pi}{2\kappa}} & e^{\frac{\om \pi}{2\kappa}} \emat. \label{SNHR}
\ee
Unlike the other matrices in \eq{Sfactogene}, $S_{\rm NHR}$
is \emph{universal} in that it only depends on $\kappa$ in \eq{kappadef}.
It is independent of the other properties of the profile $v(x)$, and also of the mass $m$.
In fact, when considering a two dimensional massless field, which obeys \eq{modequ} with $m = 0$,
the left moving v-modes decouple, $n^{\rm v}_\om$ in \eq{occupnum+} vanishes, and
the total $S$-matrix reduces to the above $S_{\rm NHR}$
(when the asymptotic flow velocity $v$ is such that {\it out} modes are well defined).
In that case, on both sides of the horizon, the flux of $u$-quanta is Planckian, and
at the standard Hawking temperature $\kappa/2\pi$, since $n^{\rm u}_\om= n^{\rm u}_{-\om}$ and
\be
\frac{n^{\rm u}_\om}{n^{\rm u}_\om + 1} = \left| \frac{\beta_\om^{\rm NHR}}{\alpha_\om^{\rm NHR}}\right|^2 = e^{- 2\pi \om/\kappa} .\label{SNHR+1}
\ee

\subsection{Exterior and interior scatterings}
\label{StrucSec}

As mentioned above, $S_{\rm ext}$ and $S_{\rm far}$ both depend on other properties of the profile $v(x)$ than $\kappa$.
However their structure can be analyzed in general terms,
and the meaning of their coefficients can be identified.
Before computing these coefficients in specific flows,
it is of value to present their general features.

Below $\om_R$ of \eq{thres}, the positive norm mode is totally reflected
while the negative norm mode propagates inside the horizon.
The exterior scattering matrix $S_{\rm ext}$
is thus fully characterized by the phase accumulated by the positive norm mode
in the right region
\be
\bmat \phi_\om^{\rm Right} \\ \left(\phi_{-\om}^{\rm Left}\right)^* \emat = \bmat e^{i\delta_\om} & 0 \\ 0 & 1 \emat \cdot \bmat \phi_\om^{\rm Refl} \\ \left(\phi_{-\om}^{\rm Left}\right)^* \emat. \label{Sextbelow}
\ee
Since this mode is totally reflected, it is the unique ABM of \eq{canmodequ} in R.
For small $x\to 0^+$, using \eq{rightm}, its behavior will be
\be
\varphi_\om(x) \underset{0^+}
{\sim} \, C \left[  |2\kappa x|^{i \frac{\om}{2\kappa} +\frac12} + e^{i\delta_\om} \times
|2\kappa x|^{-i \frac{\om}{2\kappa} +\frac12} \right], \label{Boom}
\ee
which will allow us in the next Sections to extract the phase $e^{i\delta_\om}$.

In the interior region, there is some extra mode mixing which is described by $S_{\rm far}$.
Because the norms of the two modes are of opposite sign, this mixing
introduces new Bogoliubov coefficients:
\be
\bmat \phi_\om^{\rm Left} \\ \left(\phi_{-\om}^{\rm Left}\right)^* \emat = \bmat \alpha_\om^{\rm far} & \beta_\om^{\rm far*} \\ \beta_\om^{\rm far} & \alpha_\om^{\rm far*} \emat \cdot \bmat \phi_\om^{\rm out} \\ \left(\phi_{-\om}^{\rm out}\right)^* \emat \label{Sfar}.
\ee
This scattering is entirely governed by \eq{canmodequ} in the interior region L.
When working in an appropriate basis, namely when positive and negative frequency modes are complex conjugated,
the real character of \eq{canmodequ} guarantees that the matrix $S_{\rm far}$ is an element of $SU(1,1)$.

\begin{figure}[!h]
\begin{center}
\includegraphics[scale=1]{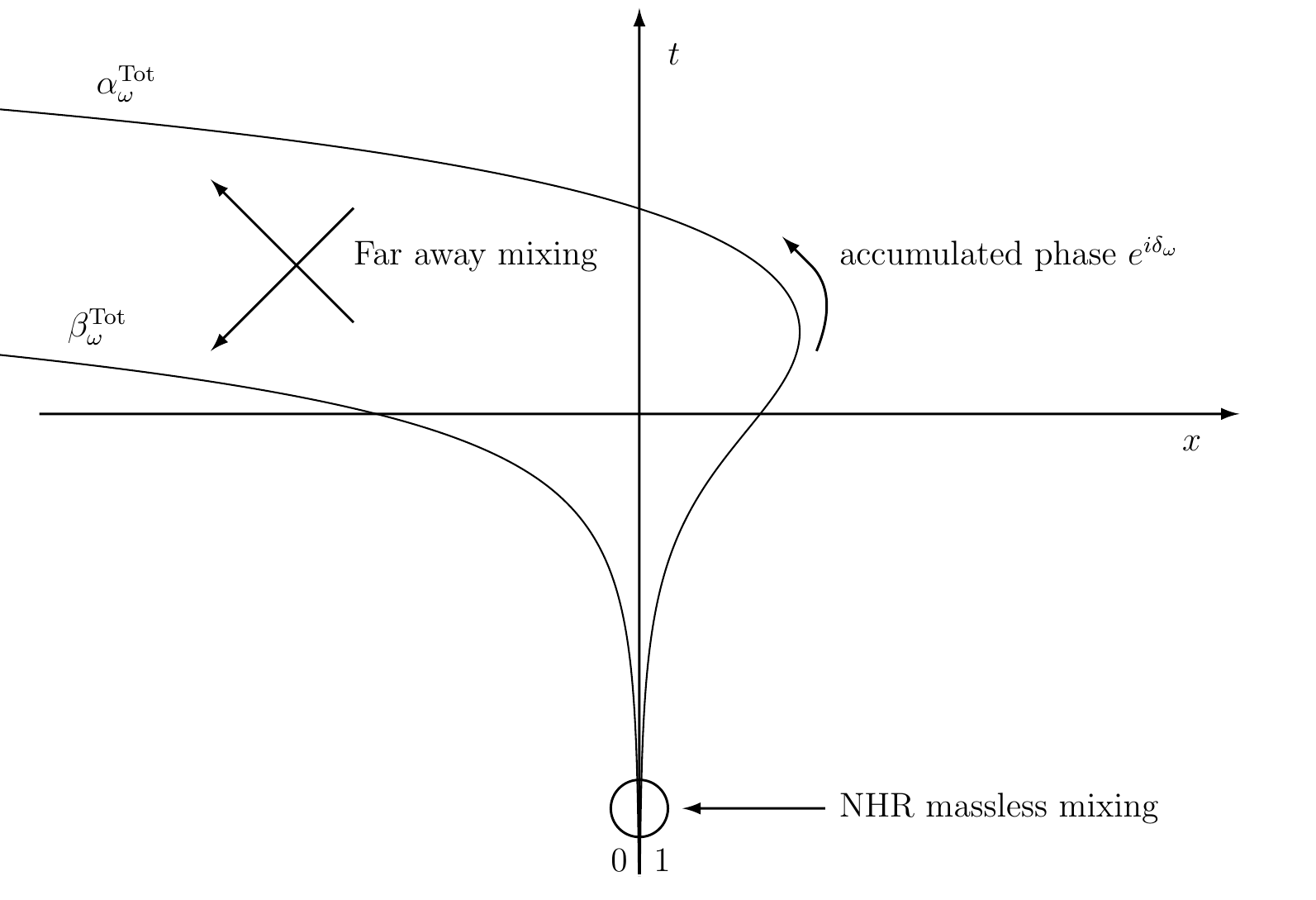}
\end{center}
\caption{In this figure a schematic representation of the $S$-matrix decomposition in \eq{Sfacto22} is shown.
The asymptotic {\it in} and {\it out} amplitudes of the positive
frequency mode $\phi_\om^{\rm in}$ are indicated.}
\label{Sfacto_fig}
\end{figure}

We see that the regularity condition of \eq{regularcondition} for the mode crossing the horizon plus the ABM requirement
reduces the dimensionality of the four unknown coefficients of \eq{NHRbehav} to two.
The $S$-matrix in \eq{Sfactogene} is then
\be
S = \bmat \alpha_\om^{\rm Tot} & \beta_\om^{\rm Tot} \\ \tilde \beta_\om^{\rm Tot} & \tilde \alpha_\om^{\rm Tot} \emat = \bmat \alpha_\om^{\rm far} & \beta_\om^{\rm far} \\ \beta_\om^{\rm far*} & \alpha_\om^{\rm far*} \emat
\cdot \bmat e^{i\delta_\om} & 0 \\ 0 & 1 \emat \cdot  \bmat \alpha_\om^{\rm NHR} & \beta_\om^{\rm NHR} \\ \tilde \beta_\om^{\rm NHR} & \tilde \alpha_\om^{\rm NHR} \emat.\label{Sfacto22}
\ee
This decomposition is depicted in Fig.\ref{Sfacto_fig}.
Notice that  it is the transposed version of
the intermediate $S$-matrices of Eqs.~\eqref{SNHR_1}, \eqref{Sextbelow}, 
\eqref{Sfar} that appear in \eq{Sfacto22}, as explained after \eq{Sdef22}.

For frequencies larger than $\om_R$, the $u$ mode and the $v$ mode, both of positive norm, mix in the exterior region R.
The matrix $S_{\rm ext}$ now is $3\times 3$ and reads
\be
\bmat \phi_\om^{\rm Right} \\ \left(\phi_{-\om}^{\rm Left}\right)^* \\ \phi_\om^{\rm in, v} \emat = \bmat T_\om & 0 & \tilde R_\om \\ 0 & 1 & 0 \\ R_\om & 0 & \tilde T_\om \emat \cdot \bmat \phi_\om^{\rm out, u} \\ \left(\phi_{-\om}^{\rm Left}\right)^* \\ \phi_\om^{\rm NHR, v} \emat.
\ee
The $2\times 2$ non trivial sector of this matrix is an element of $U(2)$,
and it describes an \emph{elastic} scattering.
The interior scattering shares the same properties as occur for frequencies below $\om_R$, namely the positive and negative norm modes propagating in the region L mix.
The outgoing $u$ mode is not affected by this scattering,
and is thus left unchanged by $S_{\rm far}$. Therefore, the structure of the total $S$-matrix is
\be
S = \bmat 1 & 0 & 0 \\ 0 & \alpha_\om^{\rm far*} & \beta_\om^{\rm far*} \\ 0 & \beta_\om^{\rm far} & \alpha_\om^{\rm far} \emat \cdot \bmat T_\om & 0 & R_\om \\ 0 & 1 & 0 \\ \tilde R_\om & 0 & \tilde T_\om \emat \cdot \bmat \alpha_\om^{\rm NHR} & \beta_\om^{\rm NHR} & 0 \\ \tilde \beta_\om^{\rm NHR} & \tilde \alpha_\om^{\rm NHR} & 0 \\ 0 & 0 & 1 \emat .\label{Sfacto33}
\ee

\section{Exactly solvable models}
\label{quattro}

We shall first compute the above $S$-matrices in three preparatory cases
in order to understand various aspects regarding the
scattering of massive modes on a stationary horizon.
The results of these three cases will then be put together so as
to obtain the $S$-matrix in a background flow
relevant for analogue gravity models,
and similar to that presented in Fig.~\ref{vprofile_fig}.

In what follows,
the various geometries will be characterized by a single function, given by the conformal factor of \eq{PGds}
\be
C(x) = 1-v^2(x). \label{conformalF}
\ee
The reason to refer only to this function is double. First, as we see from \eq{canmodequ}, the mode equation for $\varphi_\om$ only depends on $C(x)$.
Second, it will allow us to consider ranges of $x$ where $C(x) >1$.
In such regions, the function $v(x)$ of \eq{PGds}
is complex. However, neither the geometry nor the wave equation \eqref{canmodequ} is ill defined,
as can be seen by making the change of time coordinate $t_S = t + \int v dx/(1-v^2)$, which gives
\be
ds^2 = (1-v^2) dt_S^2 - \frac{dx^2}{1-v^2}.
\ee
This line element depends only on $C(x)$, and can be extended to $C>1$. In fact, the $v$ dependent prefactor in \eq{def_varphi} 
accounts for the coordinate change $t \to t_S$. The status of the auxiliary mode $\varphi_\om$
is thus clear: it is the stationary mode when using the `Schwarzschild' time coordinate $t_S$,
which is singular on the horizon.
While the auxiliary mode $\varphi_\om$ obeys an equation which is simpler to solve,
it is singular across the horizon, see \eq{rightm}. Hence we shall use the original mode $\phi_\om$ to impose regularity conditions on the horizon, see \eq{regularcondition}.

\subsection{Rindler horizon}
\label{RindlerSec}

It is instructive to first study a Rindler (future) horizon in the above formalism.
To do so, we use the profile defined by
\be
C(x)= 2\kappa x .
\label{Rindler}
\ee
It is straightforward to check that this metric has a vanishing scalar curvature
$R = - \partial_x^2 C/2$, and thus describes flat space. In \eq{Rindler} $\kappa$ is the `surface gravity'
as defined by \eq{kappadef}.
The fact that it depends on the arbitrary normalization of the Killing field $K_t$
is free of physical consequence, because the $S$-matrix depends only
on the ratio $\om/\kappa$, see {\it e.g.}, \eq{SNHR}.

In this geometry, \eq{canmodequ} reads
\be
\left[ - \partial_x^2 +\frac{m^2}{2\kappa x} -\left( \frac14 + \frac{\om^2}{4\kappa^2}\right)\frac1{x^2} \right] \varphi_\om(x) = 0 .
\label{modequRindler}
\ee
The interesting aspect of Rindler space
is that we know the result in advance.
Indeed, since there is no pair creation in flat space, the total Bogoliubov transformation of \eq{Sfactogene} must be trivial, {\it i.e.}, $\beta_\om^{\rm Tot} = 0$. However, from \eq{modequRindler} we see that close to the horizon
the modes behave as in \eq{NHRbehav},
and thus are subjected to the NHR mixing described in Sec.\ref{NHRSec}.
Therefore, the extra scattering described by $S_{\rm far}$ and $S_{\rm ext}$ exactly compensates the NHR one, so that the total $S$-matrix is trivial. 
To show that this is the case, we solve \eq{modequRindler}
following the steps of Sec.~\ref{NHRSec} and \ref{StrucSec}.

\eq{modequRindler} should thus be solved separately for $x>0$ and $x<0$.
On both sides, its solutions can be expressed in terms of Bessel functions. We start by studying the
exterior R region. For $x > 0$, the only ABM is
\be
\varphi_\om(x) = C \frac{2i}{\pi} K_{i \om/\kappa}\left(2\sqrt{\frac{m^2 x}{2\kappa}}\right),
\ee
where $K_{\nu}(z)$ is the Mac-Donald function~\cite{Olver} and $C$ a constant.
At large values of $x$ 
\be
\varphi_\om(x) \underset{+\infty}{\sim} 2i C \left(\frac{\kappa x}{8\pi^2 m^2}\right)^{\frac14} e^{-2\sqrt{\frac{m^2 x}{2\kappa}}}.
\ee
This exponential decrease is expected since $\om_R$ of \eq{thres} is infinite. Near the horizon, for $x \to 0^+$, the ABM behaves as
\be
\varphi_\om(x) \underset{0^+}{\sim} - Ce^{i\frac{\om}{\kappa} \ln(\frac m{2\kappa})} \left(\frac{e^{i\delta_{\rm Rindler}}\times |2\kappa x|^{-i\frac\om{2\kappa} +\frac12} + |2\kappa x|^{i\frac\om{2\kappa} +\frac12}}{\Gamma(1+i\om/\kappa) \sinh(\frac{\om\pi}{\kappa})}\right),
\ee
where
\be
e^{i\delta_{\rm Rindler}} = \frac{\Gamma(i\om/\kappa)}{\Gamma(-i\om/\kappa)} e^{-2i\frac\om\kappa \ln(\frac m{2\kappa})}.
\label{deltaR}
\ee
This is the phase shift that enters in \eq{Sextbelow}.
It will play a crucial role in what follows.

In the interior region L, the general solution reads
\be
\varphi_\om(x) = A \sqrt{-x} J_{-i\om/\kappa}\left(2\sqrt{-\frac{m^2 x}{2\kappa}}\right) + B \sqrt{-x} J_{i\om/\kappa}\left(2\sqrt{-\frac{m^2 x}{2\kappa}}\right).
\ee
Near the horizon, for $x \to 0^-$, one finds
\be
\varphi_\om(x) \underset{0^-}{\sim} A_\om \frac{e^{-i\frac\om{\kappa} \ln(\frac m{2\kappa})}}{\sqrt{2\kappa}\Gamma(1-i\om/\kappa)} |2\kappa x|^{-i\frac\om{2\kappa} + \frac12} + B_\om \frac{e^{i\frac\om{\kappa} \ln(\frac m{2\kappa})}}{\sqrt{2\kappa}\Gamma(1+i\om/\kappa)} |2\kappa x|^{i\frac\om{2\kappa} + \frac12}.
\ee
In order to build the normalized positive frequency left mode $\phi_\om^{\rm Left}$ appearing in \eq{Sfar},
we choose
\be
A_\om = -i\sqrt{\frac{\om}{2\pi \kappa}} \Gamma(-i\om/\kappa) e^{i\frac\om{2\kappa} \ln(\frac m{2\kappa})},
\ee
and $B_\om=0$, so as to get
\be
\varphi_\om^{\rm Left}(x) \underset{0^-}{\sim} \frac{ |2\kappa x|^{-i\frac\om{2\kappa} + \frac12}}{\sqrt{4\pi \om}}.
\ee
For $x \to - \infty$,  the asymptotic behavior of this mode is~\cite{Olver,AbramoSteg},
\begin{align}
\varphi_\om^{\rm Left} \,  \underset{-\infty}{\sim} & \sqrt{\frac{\om}{2\pi \kappa}} \Gamma(-i\om/\kappa)\, e^{\frac{\om\pi}{2\kappa}} \, e^{i\frac\om{\kappa} \ln(\frac m{2\kappa}) -i\frac\pi4} \times \left\{ \frac{e^{-i\sqrt{\frac{-2m^2 x}{\kappa}}}}{\left(\frac{8\pi^2 m^2}{- \kappa x}\right)^{\frac14}}
+ e^{-\frac{\om\pi}{\kappa}} e^{ i\frac{\pi}2} \times \frac{e^{i\sqrt{\frac{-2m^2 x}{\kappa}}}} {\left(\frac{8\pi^2 m^2}{- \kappa x}\right)^{\frac14}} \right\} .
\end{align}
In the parenthesis, the first term is the asymptotic positive norm {\it out} mode,
whereas the last factor of the second term
gives the negative norm one. Therefore, the
coefficients of \eq{Sfar} are
\begin{align}
\alpha_\om^{\rm far} &= \sqrt{\frac{\om}{2\pi \kappa}} \Gamma(-i\om/\kappa)\, e^{\frac{\om\pi}{2\kappa}} \, e^{i\frac\om{\kappa} \ln(\frac m{2\kappa}) -i\frac\pi4},\\
\beta_\om^{\rm far*} &= \alpha_\om^{\rm far} \times e^{-\frac{\om\pi}{\kappa}} \, e^{i \frac{\pi}2}.
\label{farR}\end{align}

Making a similar computation for the negative left mode $(\phi_\om^{\rm Left})^*$, we obtain $\alpha_\om^{\rm far*}$ and $\beta_\om^{\rm far}$ and verify that
$S_{\rm far}$ is an element of $SU(1,1)$.
From \eq{farR} we see that $\vert \beta_\om^{\rm far}/\alpha_\om^{\rm far}\vert^2 = e^{- 2 \pi \om/\kappa}$, which is exactly the ratio of the near horizon coefficients of \eq{SNHR}.
This is a necessary condition for having $\beta_\om^{\rm Tot} = 0$. However it is not sufficient, as one also needs the
phases to match each other, since
\be
\beta_\om^{\rm Tot} =  \tilde \alpha_\om^{\rm NHR} \beta_\om^{\rm far} + \alpha_\om^{\rm far} \beta_\om^{\rm NHR} e^{i \delta_{\om}}. \label{betaTotRind}
\ee
From this equation,
one clearly sees the crucial role played by $e^{i\delta_\om}$ of \eq{deltaR}.
An explicit calculation shows that the total $S$-matrix of \eq{Sfacto22} is
\be
S = \frac{\Gamma(i\om/\kappa)}{\Gamma(-i\om/\kappa)} e^{-i\frac\om\kappa \ln(\frac m{2\kappa})} \bmat e^{-i\frac\pi4} & 0 \\ 0 & e^{i\frac\pi4} \emat.
\ee
We see that the two $in/out$ coefficients $\beta_\om^{\rm Tot}$ vanish for all values of $\om$ and $m$.
Hence the scattering away from the horizon {\it exactly} compensates the near horizon mixing
and there is no pair creation.
Of course, this exact cancellation was expected in the present case.
However, in more general space-times, as we shall see below,
a partial cancellation between $S_{\rm far}$ and $S_{\rm NHR}$
will be obtained for similar reasons.

\subsection{Totally reflecting model}
\label{SandroSec}

Our second example generalizes the former Rindler case in that
there is still a total reflection,
but the profile $v(x)$ now possesses an asymptotically flat interior region.
As a result, the asymptotic flux of left going particles is well defined,
since the emitted quanta are asymptotically described by plane waves.
The profile which generalizes \eq{Rindler} is
\be
C(x) = D(-1 + e^{2\kappa x\over D}). \label{Sandrov}
\ee
The parameter $D$ characterizes the transition from the Near Horizon Region to the asymptotic one.
In the limit $D\to \infty$, $C(x)$ of \eq{Sandrov}
becomes that of \eq{Rindler}, which describes Rindler space.
In the above metric, \eq{canmodequ} is analytically solvable in terms of hypergeometric functions~\cite{Olver,AbramoSteg}.
The full expression of the general solution is given in App.~\ref{HypApp}.
To compute the total $S$-matrix, we follow
exactly the same procedure as in Sec.~\ref{RindlerSec}.

The first important quantity is the phase shift of \eq{Sextbelow}.
To simplify its  expression, we introduce the dimensionless quantities
\begin{align}
\varpi & \doteq \frac{\om}{2\kappa},\quad
\bar \Om_+  \doteq \frac{\sqrt{m^2D + \om^2}}{2\kappa}. \label{notations}
\end{align}
The (exact) phase shift then reads
\be
e^{i\delta_{\rm Refl}} = \frac{\Gamma(2i\varpi) \, \Gamma(1 - i\varpi + i \bar \Om_+ )
\, \Gamma(1 - i\varpi - i \bar \Om_+ )}{\Gamma(-2i\varpi) \, \Gamma(1 + i\varpi - i \bar \Om_+ )\,
\Gamma(1 + i\varpi + i \bar \Om_+)} \,  e^{2i\varpi \ln(D)} .\label{Sandrophase}
\ee
In the interior L region, the scattering coefficients in $S_{\rm far}$ are
\begin{align}
\alpha_\om^{\rm far} &= \left(\frac{\bar \Om_+^2}{\varpi^2}\right)^{\frac14} \frac{\Gamma(1-2i\varpi)\Gamma \left(-2i\bar \Om_+\right)}{\Gamma \left(-i\bar \Om_+ - i\varpi \right) \Gamma \left(1 - i\varpi - i\bar \Om_+\right)} e^{i\varpi \ln(D)} \nonumber ,\\
\beta_\om^{\rm far} &= \alpha_{-\om}^{\rm far} \times e^{-2i\varpi \ln(D)}. 
\label{Sandrophase2}
\end{align}
The total beta coefficient is then given by
\be
\beta_\om^{\rm Tot} = \tilde \alpha_\om^{\rm NHR} \beta_\om^{\rm far} + \alpha_\om^{\rm far} \beta_\om^{\rm NHR} e^{i \delta_{\rm Refl}}. \label{betaTot}
\ee
Its full expression is rather complicated, and not very transparent. It is more interesting
to study its behavior in different regimes of the parameter space $(\om/\kappa,m/\kappa,D)$.

\subsubsection{Low frequency regime}
\begin{figure}[!h]
\begin{center}
\includegraphics[scale=0.55]{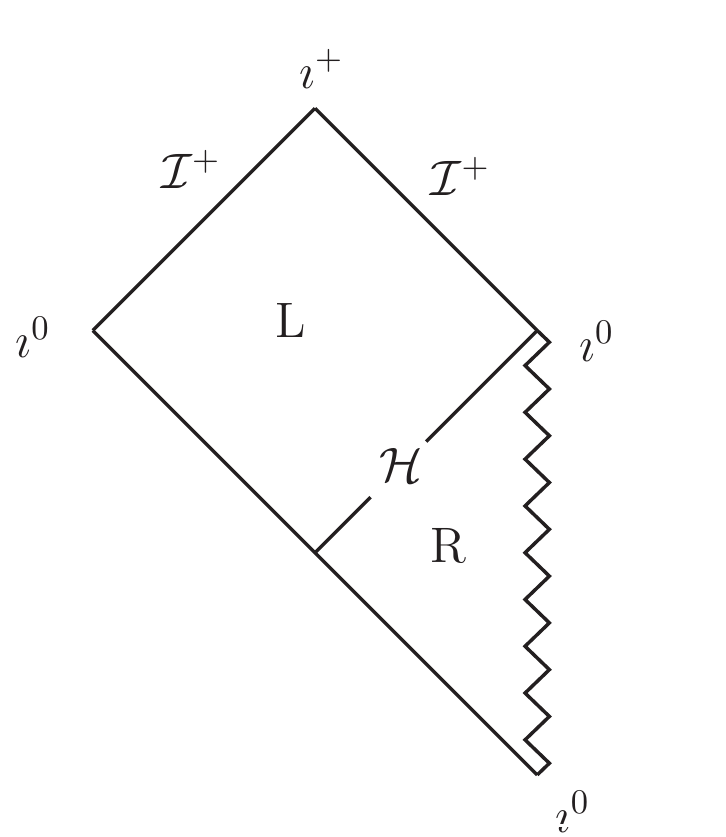}
\includegraphics[scale=0.55]{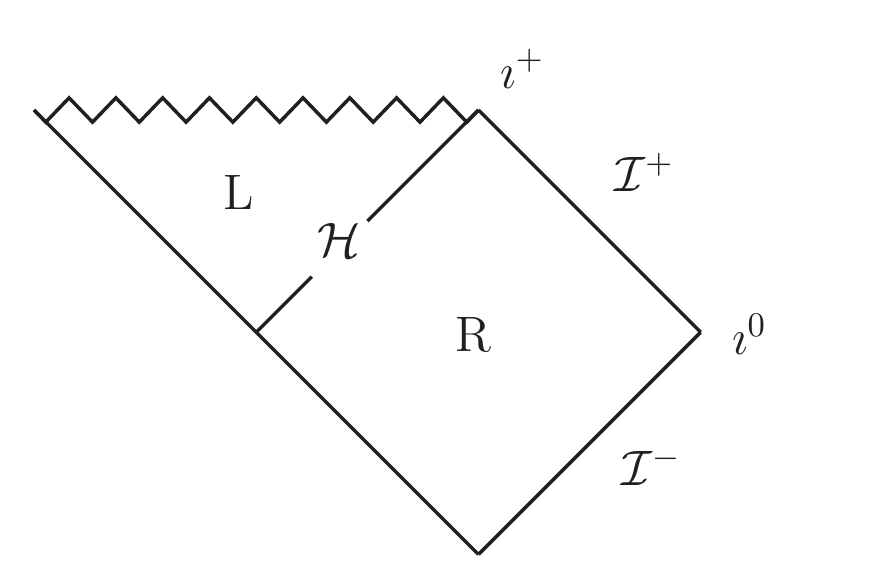}
\includegraphics[scale=0.55]{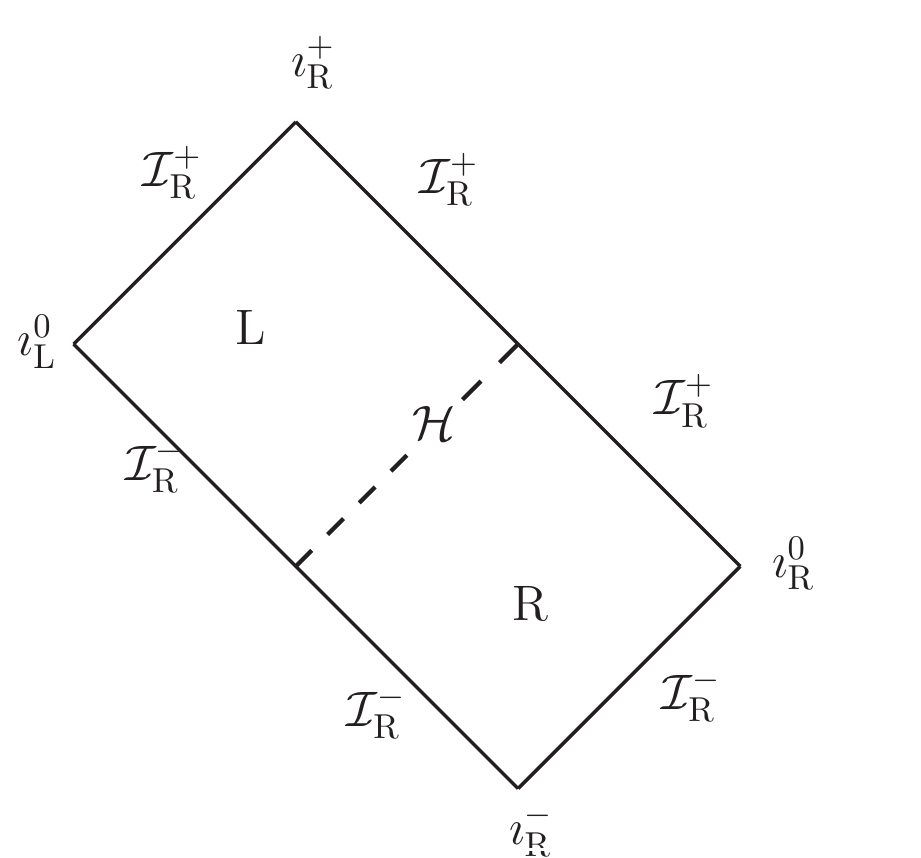}
\end{center}
\caption{Shown in this figure are the Penrose-Carter diagrams of the geometries we shall use. On the left, the totally reflecting model of \eq{Sandrov} which is singular in the exterior region, in the middle, the CGHS model of \eq{CGHSv} with its interior singularity,
and on the right, the analog model of \eq{analogv} which is everywhere regular, and can be obtained by
pasting the L quadrant of the first model with the R quadrant of the second.
These diagrams do not represent
the full analytic extension of each space-time, but only the quadrants that are relevant for our $S$-matrix,
namely, the left (L) and Right (R) regions on either
side of the (future) horizon $\mathcal H$.
Precise definitions of the various types of infinities along with more details about the 
last diagram, are given in Ref.~\cite{Barcelo04}.}
\label{3Penrose_fig}
\end{figure}

An interesting phenomenon happens in the deep infrared regime, $\om\to 0$. In this regime, we find
\begin{align}
e^{i\delta_{\rm Refl}} &\sim -1,\\
\beta_\om^{\rm far} \sim \alpha_\om^{\rm far} &
\sim \sqrt{\frac{m D^{\frac12}}\om} \frac{\Gamma \left(-i\frac{mD^{\frac12}}\kappa \right)}{\Gamma \left(-i\frac{mD^{\frac12}}{2\kappa} \right)\Gamma \left(1-i\frac{mD^{\frac12}}{2\kappa} \right)},\\
\beta_\om^{\rm NHR} \sim \alpha_\om^{\rm NHR} &\sim -i \sqrt{\frac\kappa{2\pi\om}} .
\end{align}
These equations show that, while both $\beta_\om^{\rm NHR}$ and $\beta_\om^{\rm far}$ diverge as $1/\om^{1/2}$, 
the total coefficient in \eq{betaTot} does not diverges as $1/\om$, 
as one might have expected. 
Using the analytic character of the $\Gamma$ functions in Eqs. (\ref{Sandrophase},\ref{Sandrophase2}), 
one finds that the leading term is constant, that is 
\be
\beta_{\om}^{\rm Tot} \underset{\om\to 0}\sim f(\kappa/mD^{1/2}) \label{betazero}, 
\ee
which is finite for all $m > 0$. This completely differs from
the massless case where $\beta_\om^{\rm Tot}$ diverges 
$\sim \beta_\om^{\rm NHR} \sim 1/\sqrt{\om}$, see the discussion after \eq{SNHR}.
Moreover, when we take the massless limit of \eq{betaTot} at fixed $\om$,
we obtain the massless result,
\be
\beta_\om^{\rm Tot} \underset{m \to 0}{\sim} \beta_\om^{\rm NHR}, \label{betaHawk}
\ee
for all $\om$, and thus in particular we recover the diverging behavior for $\om\to 0^+$.
Before addressing the apparent contradiction between \eq{betazero} and \eq{betaHawk},
it is of value to make a pause and to discuss the lesson from \eq{betaHawk}.
This equation shows that when a massless conformally coupled field is scattered
on a Killing horizon of a stationary metric which is asymptotically singular in the
external region (since the curvature $R= -\partial_x^2 C/2 \to \infty$ for $x\to \infty$, see Fig.~\ref{3Penrose_fig}),
the particle flux is nevertheless well defined in the {\it interior}
 region because it is asymptotically flat,
so that the $out$ modes of {\it negative} Killing frequency are unambiguously defined.
In this case, using \eq{SNHR+1}, one gets a Planck spectrum emitted towards asymptotic left infinity.  This is rather unusual since the
Killing frequency is negative, and has in that L region the physical meaning of a momentum
since the Killing field is space-like.

\begin{figure}[!h]
\begin{center}
\includegraphics[scale=1]{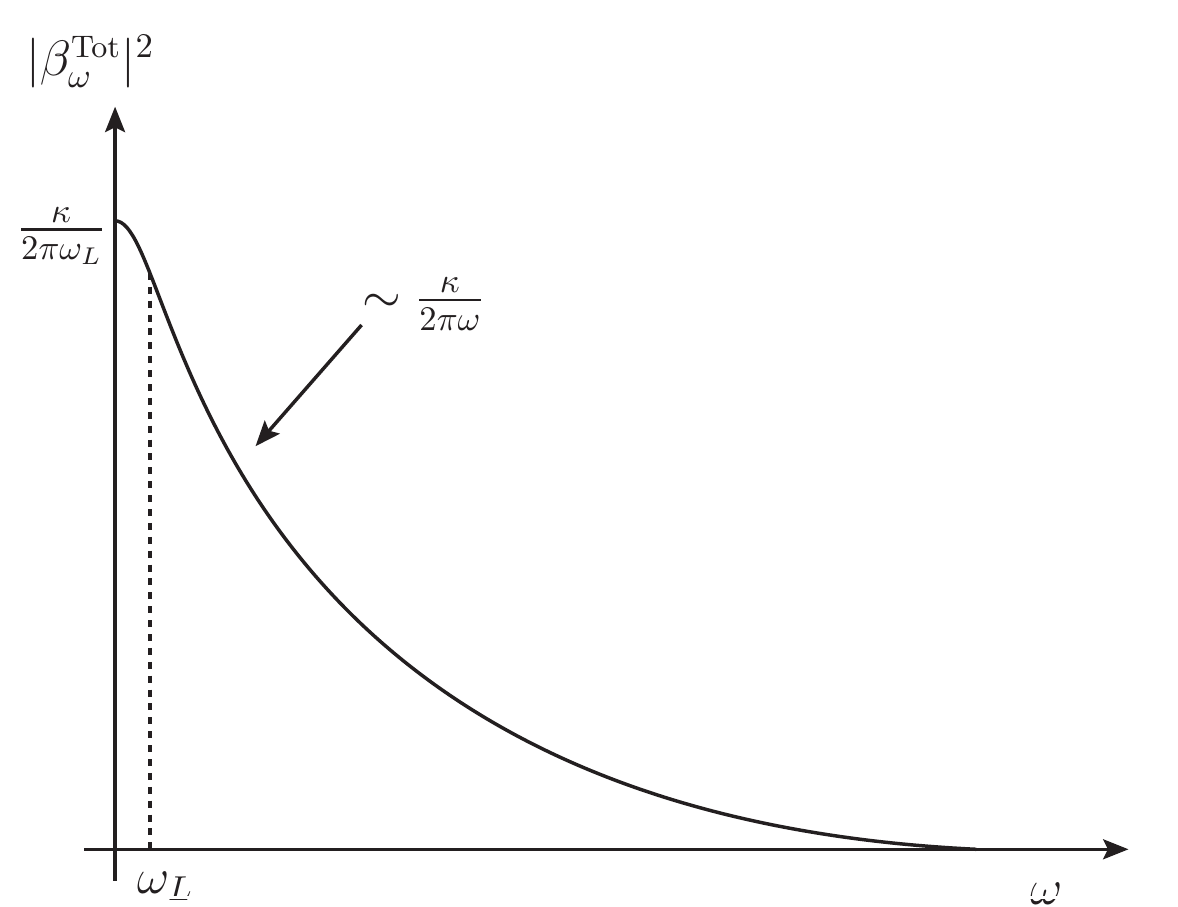}
\end{center}
\caption{In this figure $\vert \beta_\om^{\rm Tot}\vert^2$ is plotted as a function of $\om$ in the regime
of low mass and frequencies, {\it i.e.}, $\om, m \ll \kappa$. For frequencies above the threshold of \eq{omL}, $\vert \beta_\om^{\rm Tot}\vert^2$
behaves as for a massless field, and grows as $\kappa/\om$. For frequencies $\om < \om_L $,
 $\vert \beta_\om^{\rm Tot}\vert^2$ saturates at $\sim \kappa/\om_L \sim  \kappa/m D^{1/2}$.
In the opposite regime, where the mass is larger than $\kappa/2\pi$, the suppression arises at a frequency larger than the
temperature, and $\vert \beta_\om^{\rm Tot}\vert^2$ remains smaller than 1.}
\label{Sat_fig}
\end{figure}

The compatibility between \eq{betazero} and \eq{betaHawk} is understood
when realizing that the saturated value $\beta_0^{\rm Tot}$ of \eq{betazero}
diverges when $m\to 0$. To see this more precisely, we focus on the regime of small mass $m\ll \kappa$ and small frequencies $\om \ll \kappa$, for
arbitrary ratios $\om/m$. In this regime, we get
\be
\beta_\om^{\rm Tot} \sim -i \left(\frac{\kappa^2}{4\pi^2 (\om^2 + m^2 D)}\right)^{\frac14}. \label{betatransit}
\ee
This expression reveals that there is a change of regime near
\be
\om_L = m D^{\frac12}. \label{omL}
\ee
When $\kappa \gg \om \gg \om_L$, $\beta_\om^{\rm Tot}$ is growing as in the massless case,
 whereas for $\om \ll \om_L$ it saturates at a high but finite value, as can be seen Fig.~\ref{Sat_fig}.
As in the case of ultra-violet dispersion~\cite{MacherWH,ACRPSF}, we observe that the effective
frequency that governs the spectrum depends, as expected, on the dispersive frequency, here the mass $m$,
there the ultra-violet scale $\Lambda$, but also depends in a non-trivial manner on the parameter $D$ that governs the
extension of the NHR. In the present case the power of $D$ is $1/2$, whereas for ultra-violet
dispersion, the power is $(n+1)/n$ when
the dispersion relation that replaces \eq{HJ} is $\Om^2 = p^2 + p^{n+2}/\Lambda^{n}$.

\subsubsection{Large mass regime}

When the mass is large, {\it i.e.}, $m\gg \kappa, \om$, we find
that the coefficients of $S_{\rm NHR}$ and $S_{\rm far}$
go to their Rindler values in a well controlled manner, {\it e.g.},
\be
e^{i\delta_{\rm Refl}} \underset{m \to \infty}\sim e^{i\delta_{\rm Rindler}} \left(1+O(\kappa/mD^{\frac12}) \right).\label{SandrophaselimRindler}
\ee
This implies that $\beta_\om^{\rm Tot} \to 0$ for $m \to \infty$ as
\be
\beta_\om^{\rm Tot} = O(\kappa/mD^{\frac12}) = O(\kappa/\om_L).\label{SandrolimRindler}
\ee
This can be understood by considering the Bessel functions of Sec.~\ref{RindlerSec}.
Their behavior reveals that
the scattering away from the horizon in region $L$, which compensates the NHR mixing, occurs on a
distance from the horizon $\sim \kappa/m^2$. Hence, for large masses, the entire scattering
occurs in a close vicinity of the horizon.
Therefore, in the large mass limit, the scattering in the present geometry
is indistinguishable from that occurring in Rindler space.
Using WKB techniques~\cite{ACRPSF},
which furnish reliable approximations in the large mass limit,
one can demonstrate that the residual scattering outside the NHR is negligible.
This means that for $m\gg \kappa, \om$,
irrespectively of the properties of the (smooth) profile $v(x)$,
the net $in/out$ Bogoliubov coefficient $\beta_\om^{\rm Tot}$ is
suppressed by the mass. This behavior radically differs from that of the
massless case given in \eq{betaHawk}, even though both cases share the same $S_{\rm NHR}$.

\subsection{CGHS model}
\label{CGHSSec}

We now study another exactly soluble example, which is given by the CGHS black hole~\cite{CGHS},
except for the definition of the surface gravity $\kappa$ which is here given by \eq{kappadef}.
 In Painlev\'e-Gullstrand coordinates, the conformal factor reads
\be
C(x) = D(1 - e^{-\frac{2\kappa x}{D}}). \label{CGHSv}
\ee
Even though this geometry is very different from that of \eq{Sandrov} as it is
singular in the interior region, see Fig.~\ref{3Penrose_fig},
at the level of the mode equation, it gives something very close since
the discrete interchange $C \to -C$ and $x\to -x$ maps one problem into the other.
For this reason, the solutions of \eq{canmodequ} will also be
hypergeometric functions, see App.~\ref{HypApp}.
As in Sec.~\ref{SandroSec}, $\kappa$ is the surface gravity and $D$ characterizes the transition from the NHR to the asymptotic region.
However, here $D$ also controls the value of the threshold frequency $\om_R$ in \eq{thres} since
\be
\label{omR}
\om_R = m D^{\frac12}.
\ee

When $\om < \om_R$, the positive norm mode is totally reflected, and  the accumulated phase shift
characterizes $S_{\rm ext}$. As in the preceding section, to obtain simpler expressions,
we introduce
\begin{align}
\bar \Om_< &= \frac{\sqrt{ \om_R^2 - \om^2}}{2\kappa},\\
\bar \Om_> &= \frac{\sqrt{\om^2 - \om_R^2}}{2\kappa},
\end{align}
which are modified versions of \eq{notations}.
The exterior phase shift is then
\be
e^{i\delta_{\rm CGHS}} = \frac{\Gamma(2i\varpi)\Gamma \left(1 - i\varpi + \bar \Om_< \right)\Gamma \left(- i\varpi + \bar \Om_< \right)}{\Gamma(-2i\varpi)\Gamma \left(1 + i\varpi + \bar \Om_< \right)\Gamma \left(i\varpi + \bar \Om_< \right)}e^{2i\varpi \ln(D)} \label{deltaCGHS}.
\ee
From this, conclusions similar to that of Sec.~\ref{SandroSec} can be drawn.
For instance, when $\om \to 0$, we recover
\be
e^{i\delta_{\rm CGHS}} \underset{\om \to 0}\sim -1 ,
\ee
which is the main ingredient needed to obtain a canceling effect as in \eq{betazero}, and to have $\beta^{\rm Tot}_{\om \to 0}$ be regular in the limit $\omega \rightarrow 0$.
If the mass is large then the behavior is essentially that found for Rindler spacetime in \eq{SandrophaselimRindler},
as can be seen by an explicit calculation.

When $\om > \om_R$, we are in the configuration where there exist three ABM, as in \eq{Sfacto33}. The greybody factors in the external region R are analytically obtained from the hypergeometric functions. The
transmission and reflection coefficients are
\begin{align}
T_\om = \tilde T_\om &= \left(\frac{\varpi^2}{\bar \Om_>^2}\right)^{\frac14} \frac{\Gamma \left(-i\bar \Om_> - i\varpi \right) \Gamma \left(1 - i\varpi - i\bar \Om_> \right)}{\Gamma(1-2i\varpi)\Gamma \left(-2i\bar \Om_> \right)} e^{-i\varpi \ln(D)} . \\
\tilde R_\om &= - \frac{ \Gamma(1+ 2i\varpi) \Gamma \left(-i\bar \Om_> - i\varpi \right) \Gamma \left(1 - i\varpi - i\bar \Om_> \right)}{\Gamma(1- 2i\varpi )\Gamma \left(-i\bar \Om_> + i\varpi \right) \Gamma \left(1 + i\varpi - i\bar \Om_> \right)} e^{-2i\varpi \ln(D)} ,\\
R_\om &= \frac{ \Gamma \left(2i\bar \Om_>\right) \Gamma \left(-i\bar \Om_> - i\varpi\right) \Gamma \left(1 - i\varpi - i\bar \Om_>\right)}{ \Gamma \left(-2i\bar \Om_>\right) \Gamma \left(i\bar \Om_> - i\varpi\right) \Gamma \left(1 - i\varpi + i\bar \Om_>\right)} .
\end{align}
Using \eq{Sfacto33}, the asymptotic out-going flux of \eq{occupnum+} is
\be
n_\om^{\rm u} = \langle 0, in| (a_\om^{{\rm out, u}})^\dagger a_\om^{{\rm out, u}}|0, in \rangle = |\beta_\om^{\rm NHR} T_\om|^2
\ee
At $\om =\om_R$, $T_\om$ vanishes, and below $\om_R$ it is trivially 0.
In the next section, the transition shall be analyzed in more detail.

\subsection{Analog model}
\label{acousticSec}

We now consider a profile that combines the regular interior region of Sec.~\ref{SandroSec}
with  the regular exterior region of the above CGHS model so as to get a flow
similar to that of Fig.~\ref{vprofile_fig}.
The resulting geometry is relevant for analog models where the velocity profile is everywhere bounded.
We thus consider
\be
\label{analogv}
C(x) = 1-v^2(x) = \left\{ \begin{aligned}
&D_L(-1 + e^{\frac{2\kappa x}{D_L}}) \text{ for } (x<0),\\
&D_R(1 - e^{-\frac{2\kappa x}{D_R}}) \text{ for } (x>0) .\\
\end{aligned} \right.
\ee
This profile is $C^1$, {\it i.e.}, it is continuous and its first derivative is continuous. This ensures that the global geometry obtained is {\it regular}, in the sense that the
curvature does not contain a distributional contribution~\footnote{A.C. would like to thank Y. Bardoux for fruitful discussions concerning this possibility and for suggesting reference~\cite{Poissonbook},
which discusses this type of regularity issue.}.
Since the scattering matrices $S_{\rm ext}$ and  $S_{\rm far}$ have been
already studied both in the  exterior and interior regions,
all we need to do here is to combine them to get the total $S$-matrix
\be
S =\underbrace{S_{\rm far}}_{\text{Sec.\ref{SandroSec}}}\cdot \underbrace{S_{\rm ext}}_{{\text{Sec.\ref{CGHSSec}}}} \cdot  \,\underbrace{ S_{\rm NHR} }_{{\text{Sec.\ref{NHRSec}}}}\, . \label{Stotanalo}
\ee
The two threshold frequencies of \eq{omR} and \eq{omL} are now
\be
\label{2critfr}
\om_R = mD_R^{\frac12}, \qquad \om_L = mD_L^{\frac12}.
\ee
Having different values for $D_R$ and $D_L$ will allow us to distinguish their roles.

We first consider the totally reflecting regime, $\om<\om_R$.
Interestingly, we recover the transition seen in Sec.~\ref{SandroSec} and in Fig.~\ref{Sat_fig}.
Indeed, for $\om \ll \om_R$
\be
e^{i\delta_{\rm CGHS}} \sim -1.
\label{deltaCG}
\ee
Together with the coefficients of $S_{\rm far}$, this ensures that $\beta_\om^{\rm Tot}$ has a finite value in the limit $\om \to 0$. More precisely, in the high $\kappa$ regime, for $m, \om \ll \kappa$, we have
\be
\vert \beta_\om^{\rm Tot} \vert^2 \sim \frac{\kappa\, (D_L+D_R)}{2\pi D_R\, \sqrt{\om^2 + \om_L^2}}. \label{betalowkappa}
\ee
To observe a divergent regime $|\beta^{\rm Tot}_\om|^2 \propto \kappa/\om$,
one needs to assume that $\om_L \ll \om < \om_R$,
where the last inequality is required in order to be
below the threshold $\om_R$. This implies $D_L \ll D_R$, hence
\be
\vert \beta_\om^{\rm Tot} \vert^2 \sim \frac{\kappa}{2\pi}\frac{1}{ \sqrt{\om^2 + \om_L^2 }}. \label{betatransitL}
\ee
This expression shows the transition between the diverging regime
at the standard temperature, which is independent of $D_R$ and $D_L$,
and a saturating regime governed by $\om_L$.
For large masses $m \gg \kappa, \om$, the results are the same as in
Secs.~\ref{SandroSec} and \ref{CGHSSec}, namely the various scattering coefficients asymptote to their Rindler values.
In numerical simulations~\cite{to-appear} all of these
results have been recovered
using rather different settings where the sound speed $c$ varies with $x$ and the velocity $v$ is a constant. This demonstrates that
the low frequency behavior of \eq{deltaCG} applies to a much wider class of situations
than the one considered here.

We now have all the ingredients necessary to study the effects of a massive field
on the outgoing fluxes when starting from vacuum.
On the right side, the outgoing particle flux is that of
Sec.~\ref{CGHSSec}:  as expected, it vanishes below $\om_R$
and above it is given by
\be
n_\om^{\rm u} = |\beta_\om^{\rm NHR}\,  T_\om|^2.
\ee
To observe the transition, we work in the high $\kappa$ regime, {\it i.e.}, $\om, m \ll \kappa$, and obtain
\be
n_\om^{\rm u} \simeq \frac{\kappa}{2\pi} \Theta(\om - \om_R) \frac{ 4\sqrt{\om^2 - \om_R^2 }}{\left(\sqrt{\om^2 -\om_R^2 } + \om \right)^2}. \label{outfluxR}
\ee
The flux is thus continuous when crossing $\om_R$.

\begin{figure}[!h]
\begin{center}
\includegraphics[scale=0.6]{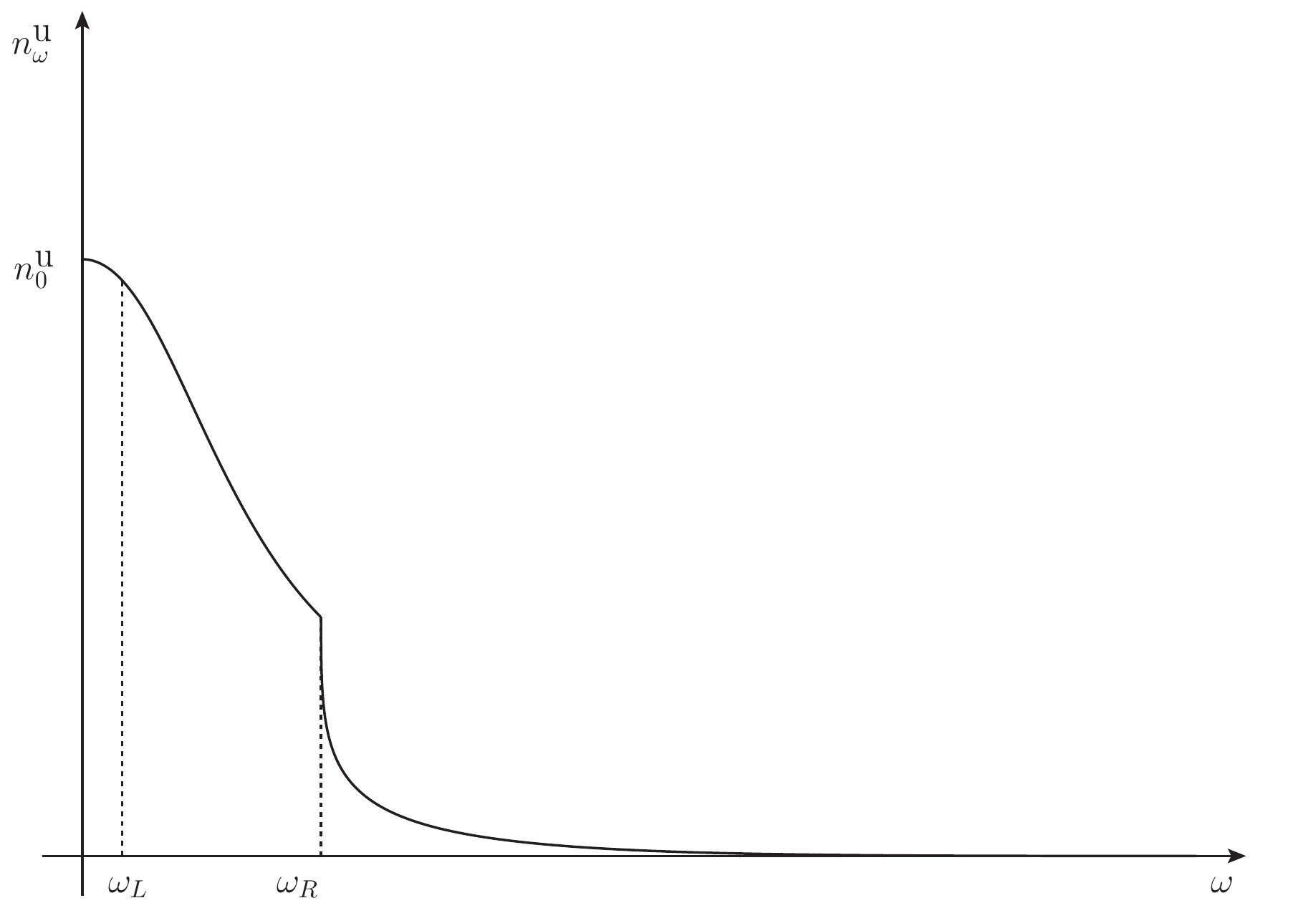}
\end{center}
\caption{Particle flux of positive frequency quanta spontaneously
emitted from the horizon towards $x \to - \infty$ in the high $\kappa$ regime.
Above $\om_R$, one finds a small contribution
which is due to the reflection (the backscattering) of Hawking quanta emitted towards $x = \infty$.
Below the threshold, the entire thermal flux is reflected and grows like $\kappa/\om$ for decreasing
values of $\om$, until one reaches $\om_L$ where it saturates, as explained in Sec.~\ref{SandroSec}.}
\label{fluxes_fig}
\end{figure}

On the left side, the particle fluxes are
more complicated since two contributions are present, see \eq{occupnum+};
$n_\om^{\rm v}$ is composed of positive frequency particles
and $n_{-\om}^{\rm u}  = n_\om^{\rm u} +  n_\om^{\rm v}$ is composed of 
the negative frequency partners. We first notice that both of these are
well defined since the profile of \eq{analogv} is asymptotically flat in L.
Using \eq{Sfacto33}, in full generality, $n_\om^{\rm v}$ reads
\be \begin{aligned}
n_\om^{\rm v} = \, &\Theta(\om - \om_R) \, |\tilde R_\om \, \alpha_\om^{\rm far}  \beta_\om^{\rm NHR} + \tilde \alpha_\om^{\rm NHR} \beta_\om^{\rm far}|^2 \\
 & +\Theta(\om_R - \om) \, | e^{i \delta_{\rm CGHS}}\, \alpha_\om^{\rm far}  \beta_\om^{\rm NHR} + \tilde \alpha_\om^{\rm NHR} \beta_\om^{\rm far} |^2.
\end{aligned} \ee
The first term in the first line, which is proportional to $\tilde R_\om$,
describes the stimulated production in the L region due to the reflected Hawking quanta.
The other terms describe the interference between the mixing in the NHR region and the scattering in the $L$ region away from the horizon.
Just as for $e^{i \delta_{\rm CGHS}}$ below the threshold, the phase of $\tilde R_\om$ is crucial since there is interference between these two terms.
Particularly interesting is behavior of $n_\om^{\rm v}$ near the threshold frequency $\om_R$.
In the regime of large surface gravity,  $\kappa \gg \om, m$,
for $\om>\om_R > \om_L$, one finds
\be
n_\om^{\rm v} \simeq \frac{\kappa}{2\pi}\frac{1}{ \sqrt{\om^2+\om^2_L}} \left(\frac{\sqrt{\om^2 + \om^2_L} - \sqrt{\om^2 - \om^2_R}}{\sqrt{\om^2 - \om^2_R} + \om}\right)^2,
\ee
whereas for $\om<\om_R$
\be
n_\om^{\rm v} \simeq \frac{\kappa }{2\pi}\frac{(1+ D_L/D_R)}{ \sqrt{\om^2 + \om^2_L}}.
\ee
These two equations describe the 
effects on the spectrum in the $L$ region
which are due to a small mass. We first notice that
$n_\om^{\rm v}$ is continuous across $\om_R$, but with a cusp, see Fig.~\ref{fluxes_fig}.
From \eq{outfluxR}, we see that this is also true for $n_\om^{\rm u}$ and hence for $n_{-\om}^{\rm u}$ as well.

We also see that
the spectrum depends on the mass $m$
only through the two critical frequencies of \eq{2critfr}. It is thus through them
that the profile properties, namely $D_R$ and $D_L$ which govern the extension of the NHR on the right and on the left, see Fig.~\ref{vprofile_fig}, affect the spectrum. The lesson here is that when dealing with a conformally invariant massless field, the surface gravity is the
only background quantity that affects the spectrum. When breaking conformal invariance,
by a mass, or a non conformal coupling as in 3+1 dimensions, or by adding some
ultraviolet dispersion, other properties of the background flow affect the spectrum.
From the above analysis, and that of~\cite{ACRPSF,MacherWH,Finazzibroad,Finazzi2regimes},
the most important ones are the extensions of the near horizon region, on both sides.

\section{Massive modes and undulations}
\label{UndulSec}

In~\cite{Mayoral2011}, it was noticed that, in a Bose condensate,
the density-density correlation function
computed in a stationary white hole flow displays an infrared divergence
with a specific short distance pattern.
As explained in the Introduction,
this divergence originates from the combination of three properties, firstly the divergence of the
Bogoliubov coefficient  $\beta_\om\sim 1/\om^{1/2}$ for $\om \to 0$,
secondly the existence of the zero frequency root $p_{\om = 0}^\Lambda\neq 0$ of
\eq{pU}, and thirdly the group velocity
oriented away from the sonic horizon.
As a result, the zero frequency phonons emitted by the white hole give a large
contribution to the  two-point function which has a short wavelength pattern.
Moreover, this contribution factorizes~\cite{ACRPSF} in
the product of twice the same real wave,
as in a presence of a coherent state of phonon modes~\cite{MacherBEC}.

In brief, the analysis of the analog Hawking radiation
in dispersive media (with either superluminal or subluminal dispersion~\cite{ACRPSF})
{\it predicts} that white hole flows should emit a zero-frequency wave
which has a large amplitude and which behaves classically.
Interestingly, this type of wave called undulation is well known in hydrodynamics~\cite{Hydro,Thesis}. 
Moreover, undulations have been observed in recent experiments aiming to detect the analogue
Hawking radiation in water tanks~\cite{Rousseaux,SilkePRL2010}.  However their intimate relationship to
Hawking radiation has not been previously pointed out.
Given the fact that the low frequency Bogoliubov coefficient $\beta_\om$
contributes to the undulation amplitude,
we think that studying and observing these waves should be conceived as part of the enterprise
to observe the analogue Hawking radiation.
As a last remark, we wish to stress that the linearized treatment predicts that, when starting from a vacuum or a thermal state,
the undulation amplitude is described by a Gaussian stochastic ensemble
with a vanishing mean amplitude,
exactly as primordial density fluctuations in the inflationary scenario~\cite{CampoRP_partiallydec}. 
Indeed, the mode mixing at the horizon amplifies vacuum or thermal fluctuations, and thus generates a random noise.
But when the initial state is classical, because it contains 
a large wave packet described by a coherent state, the outcome of the amplification is
also deterministic.

Adding a mass, or a perpendicular momentum, to a two dimensional
massless relativistic dispersion relation, also
engenders a nontrivial zero frequency root of \eq{HJ} in the supersonic L region, thereby
opening the possibility of having a `massive undulation'.
This possibility has been the main motivation of this paper.
In the following, we first present the necessary and sufficient conditions to find an undulation, starting from the two-point function, as it clearly reveals a subtle aspect,
namely the cumulative role of low frequencies
in determining the amplitude of the zero-frequency wave.
Then, we study the role of a non zero mass,
both in black and white holes flows.

\subsection{General properties}
We consider the two-point correlation function evaluated in the {\it in}-vacuum
\be
G(t,t';x,x') = \langle 0, in | \hat \phi(t,x) \hat \phi(t',x') |0, in\rangle.
\ee
Using \eq{phidecomp} we find for equal times that
\be
G(t,t;x,x') = \int_0^\infty G_\om(x,x') d\om.
\label{Gint}
\ee
In the infrared sector, for frequencies below $\om_R$ in \eq{thres},
using \eq{Sdef22}, the relevant term in the integrand is
\be
G_\om(x,x') = \phi_\om^{\rm in}(x) \left(\phi_\om^{\rm in}(x')\right)^* + \phi_{-\om}^{\rm in}(x) \left(\phi_{-\om}^{\rm in}(x')\right)^* . \label{Gom}
\ee
In general, $\phi_\om^{\rm in}$ and $\phi_{-\om}^{\rm in}$ are two different functions of $x$, and $G_\om$ is complex and cannot be factorized. 
To see that under some conditions it factorizes,
we use \eq{Sfacto22} to work with the {\it out} modes. Then the first term in Eq.~(\ref{Gom}) becomes
\begin{align}
\phi_\om^{\rm in}(x) \left(\phi_\om^{\rm in}(x')\right)^* =& |\alpha_\om^{\rm Tot}|^2\,  \phi_\om^{\rm out}(x) \left(\phi_\om^{\rm out}(x')\right)^* + |\tilde \beta_\om^{\rm Tot}|^2 \, \left(\phi_{-\om}^{\rm out}(x)\right)^* \phi_{-\om}^{\rm out}(x')
\nonumber \\
&+ 2 {\rm Re}\left\{ \alpha_\om^{\rm Tot} \tilde \beta_\om^{\rm Tot*} \phi_\om^{\rm out}(x) \phi_{-\om}^{\rm out}(x')\right\}.
\label{phiphi}
\end{align}
In the limit $\om \to 0$, two effects are combined.
First  $\phi^{\rm out}_{\om}$ and $\phi^{\rm out}_{-\om}$ become the same function of $x$,
$\phi^{\rm out}_0(x)$. This is true in general, but not
in the particular case of the two dimensional massless field in a black hole metric
because in that case $\phi^{\rm out}_{\om}$ and $\phi^{\rm out}_{-\om}$
vanish on the L and R quadrant respectively (see~\cite{ACRPSF}).

Second, when assuming
that $\vert \beta^{\rm Tot}_\om|^2  \gg 1$ for $\om \to 0$,
since $S$ of \eq{Sfacto22} is an element of $U(1,1)$,
one has
\begin{align}
|\alpha_\om^{\rm Tot}|^2 &\sim|\tilde \alpha_\om^{\rm Tot}|^2 \sim |\beta_\om^{\rm Tot}|^2 \sim |\tilde \beta_\om^{\rm Tot}|^2 , \nonumber \\
\alpha_\om^{\rm Tot} \tilde \beta_\om^{\rm Tot*} &\sim \tilde \alpha_\om^{\rm Tot*} \beta_\om^{\rm Tot} \sim e^{2i\theta}|\beta_\om^{\rm Tot}|^2,
\end{align}
where $e^{2i\theta}$ is a phase. These two facts guarantee that $G_\om$ becomes real and factorizes as 
\be
G_\om(x,x') \sim 8 |\beta_\om^{\rm Tot}|^2 \times \Phi_{\rm U}(x)\,  \Phi_{\rm U}(x') ,
\label{omundul}
\ee
where the real wave
\be
\Phi_{\rm U}(x) \doteq \Re\left\{e^{i\theta} \phi^{\rm out}_0(x) \right\},
\ee
gives the profile of the undulation. It should first be noticed that its phase is locked. Indeed, if one modifies
the arbitrary phase of the $out$ mode $\phi^{\rm out}_0$, the modified phase $\theta$
would exactly compensate this change so that $\Phi_{\rm U} $ would remain unchanged.
This can be understood from the fact that $\Phi_{\rm U} $ oscillates on one side of the horizon and decays on the other.
In fact it behaves like an Airy function, even though the phase $e^{i \theta}$ is different,
see~\cite{ACRPSF} for details.
One should point out that this factorization means that the undulation
contributes to observables in the same way that a coherent state does, see App.~C in Ref.~\cite{MacherBEC}.
It thus behaves as a classical wave in that its profile and its phase are not random.
However, in the present linearized treatment,
its amplitude is still a random variable, {\it i.e.}, its mean value
is identically zero, and $|\beta_\om^{\rm Tot}|^2$
gives the ($\om$-contribution of its) standard deviation.
This should be conceived as an important {\it prediction} of the linearized treatment,
and could be validated for a BEC using numerical techniques similar to those of~\cite{Mayoral2011},
and perhaps also in future experiments.

So far we have worked at fixed $\om$. We now consider the integral over low frequencies
in \eq{Gint}. We recall that the divergence of
$|\beta_\om^{\rm Tot}|^2$ for $\om \to 0$ accounts for a growth in time of the r.m.s. amplitude of the undulation~\cite{Mayoral2011,ACRPSF}.
When considering an observable evaluated
at a finite time $t$ after the formation of the horizon,
the stationary settings of \eq{Gint}
with a dense set of frequencies should be used with care.
Indeed, after such a lapse, one cannot resolve frequencies
separated by less than $2\pi/t$, as in the Golden Rule.
This effectively introduces an infrared cut-off in the integral over $\om$.
Taking this into account  gives the growing rate
that depends on the power of the divergence of $|\beta_\om^{\rm Tot}|^2$.
For example, when the Bogoliubov coefficients take their massless values in \eq{SNHR},
in the vacuum, the infrared contribution of $G$ grows as
\begin{align}
G_{\rm IR}(t;x,x') &\sim 8 \int_{2\pi/t} \frac{ d\om}{\om} \, \frac{ \kappa}{2\pi} \times   \Phi_{\rm U}(x)\,  \Phi_{\rm U}(x') ,
\nonumber\\
& \sim \frac{4\kappa}{\pi} \ln(t/2\pi)\times  \Phi_{\rm U}(x)\,  \Phi_{\rm U}(x') .
 \label{Gundul}
\end{align}
Of course, in a medium, this growth would saturate because of the non-linearities, as was observed
in a Bose condensate~\cite{Mayoral2011}. In the experiments of~\cite{Rousseaux,SilkePRL2010}, only a constant (saturated) amplitude was observed.
It would be very interesting to conceive experiments to observe the growth of the undulation amplitude. 
It would be also important to understand if the randomness of the amplitude
found in the linearized treatment (in vacuum and in thermal states)
is replaced by a deterministic non-linear behavior when non-linearities are included, 
or if some residual randomness persists. 
In conclusion, the linearized treatment of perturbations 
predicts that all (sufficiently regular~\cite{Finazzibroad}) white hole flows in dispersive media 
should emit an undulation with a significant amplitude.

\subsection{Massive hydrodynamical undulations in black holes}

We now study the undulation in the analog black hole metric of \eq{analogv}.
In preceding sections, we saw that the low frequency massive modes end up
in the inside L region for both signs of $\om$, see Fig.~\ref{mass_traj_fig}.
 Moreover, in the limit $\om \to 0$,
their momentum (solution of \eq{HJ}, see also Fig.\ref{WHdisprelm_fig}), is finite and given by
\be
p_U^m = p_{\om \to 0} = m D_L^{-\frac12}
\left(1+O\left(\frac{\om}{\om_{\rm U}}\right)\right).
\label{omU}\ee
This means that the zero frequency mode $\phi^{\rm out}_0(x)$ is a non trivial function of $x$,
opening the possibility of finding a behavior similar to that of \eq{Gundul}.
When $\om_L \ll \kappa$, an explicit calculation of
$ \Re\left\{e^{i\theta} \phi^{\rm out}_0(x) \right\}$,
similar to that made in \cite{ACRPSF},
tells us that the asymptotic profile of $\Phi_{\rm U}^m$ is
\be
\Phi_{\rm U}^m(x) = \frac{1}{\sqrt{4\pi \om_L}}\cos\left( p_U^m \, x \right) .
\ee

The next important aspect concerns the
calculation of the net contribution of low frequency modes to $G$.
In this respect two aspects should be discussed.
The first one concerns the fact that
$|\beta_\om^{\rm Tot}|^2$ no longer diverges for $\om\to 0$.
However, the criterion for the factorization of $G$ is only that $|\beta_\om^{\rm Tot}|^2 \gg 1$.
When $\om_L \ll \kappa$, as shown in \eq{betatransitL},
this is the case for frequencies $\om \ll \kappa/2\pi$.
The second aspect concerns the frequency interval $0\leqslant \om < \om_{\rm U}$ such that the momenta $p_\om$ are close
enough to undulation momentum $p_U$ so that the $out$ modes $\phi_\om^{\rm out}$
contribute coherently to $\phi^{\rm out}_0$.
Using \eq{HJ} in the asymptotic interior region, we get
\be
\om_{\rm U} \simeq \frac{\om_L} {\sqrt{1+D_L}} \ll \kappa. \label{omUm}
\ee
Therefore,  at time $t$ after the formation of the horizon, the
contribution of the low frequency modes to the 2 point function is
\be
\label{Gir}
G_{\rm IR}(t;x,x') = 8\int_{2\pi/t}^{\om_{\rm U}}  |\beta_\om^{\rm Tot}|^2 d\om \times \Phi_{\rm U}^m(x)\,  \Phi_{\rm U}^m(x') .
\ee
When assuming $\om_{\rm U} \ll \kappa$, which is the case for a small enough mass,
using \eq{betatransitL} we get
\be
\int_{2\pi/t}^{\om_{\rm U}}  |\beta_\om^{\rm Tot}|^2 d\om = \frac{\kappa}{2\pi} \left[ {\rm sinh}^{-1}\left(\frac{\om_{\rm U}}{\om_L} \right) - {\rm sinh}^{-1}\left(\frac{2\pi}{\om_L t}
\right) \right].
\ee
Hence, for short times, the amplitude grows as $\ln(t)$, as in the massless case.
However, when $t > 2\pi/\om_L$,
the amplitude saturates and stays constant afterwards.
This is an important prediction of this paper.
It shows how the transition in $\om$ near $\om_L$ with respect to the massless spectrum
(see Fig.~\ref{Sat_fig}) produces here a change in time of the growth rate.

To conclude this section, we wish to provide a qualitative evaluation of the importance of
this infrared contribution to $G$. To do so, we need to consider some
observables, such as the stress energy tensor.
In particular, its trace accounts for the mass density of the field
\be
{\rm Tr}(\hat T) = \langle \hat T^\mu_{\ \mu} \rangle = m^2 \langle \phi^2(x) \rangle.
\ee
At late times, {\it i.e.}, $t\gg 2\pi/\om_L$,
the contribution of the undulation to the trace is
\be
{\rm Tr}(\hat T_{\rm IR}) = \frac{\kappa m}{\pi^2 D_L^\frac12} \underbrace{\sinh^{-1}\left(\frac{\om_{\rm U}}{\om_L}\right)}_{\lesssim 1} \times
\cos^2\left({p_U^m\, x} \label{BHundul} \right).
\ee
Since we work with $\kappa \gg \om_L = m D_L^{1/2}$,
this contribution to the trace is much smaller than $\kappa^2$, the typical energy density
contained in the Hawking flux for a light field.
Hence we expect that the undulation will not be easily
visible in this case. Moreover, when $D_L\to \infty$, which is the Rindler limit, the amplitude goes to 0 as $1/D_L$,
confirming the stability of Minkowski space. It is also interesting to notice that the parameter $D_R$ plays no role (as long as $\om_{\rm} \ll \om_R$, so that \eq{betatransitL} stands),
confirming that the undulation is controlled by the interior geometry.

\subsection{Massive dispersive undulations in white holes}

For white holes, undulations can be found
when the dispersion relation is non-relativistic in the ultraviolet sector,
as discussed in the Introduction. For simplicity, we consider here the
superluminal relation $\Omega^2 = m^2 + p^2 + p^4/\Lambda^2$
which is obtained from \eq{BogdrNl} in the limit $\xi \, p_\perp \to 0$.
Notice however that both superluminal and subluminal dispersion relations
give rise to undulations in white hole flows, in virtue of the symmetry which relates them
when interchanging at the same time the R and L regions, see Sec.~III.E in~\cite{ACRPSF}.

For superluminal quartic dispersion,
the outgoing momentum at zero frequency is found in the supersonic region and is given by
\be
p_U^\Lambda = p_{\om \to 0} = \Lambda D_L^{1/2} \left( 1+O\left(\frac{\om}{\om_{\rm U}^\Lambda}\right)\right),
\ee
and the asymptotic behavior of the undulation is~\cite{ACRPSF}
\be
\Phi_{\rm U}^\Lambda (x)=\frac{\cos(p_U^\Lambda \, x + \theta_U)}{\sqrt{4\pi \om_L}}. 
\ee
The phase $\theta_U$ cannot be obtained from the preceeding equations, because it
is mainly determined by the dispersive properties of the modes.
Using the results of~\cite{ACRPSF},  one can establish that, when $m=0$, $\theta_U = (\Lambda D_L^{3/2})/(6\kappa) + \pi/4$, see also~\cite{CoutantPhD}.

In the presence of ultraviolet dispersion, the width of frequencies that contribute coherently to $G$ is
\be
\om^\Lambda_{\rm U} = \Lambda D_L^{\frac32}. \label{omUL}
\ee
Since $\Lambda$ can be much larger than $\kappa$, $\om^\Lambda_{\rm U}$ can be either smaller
or larger than the Hawking temperature $\kappa/2\pi$.
In what follows we work with $\om^\Lambda_{\rm U} \gg \kappa$
where the Bogoliubov coefficients are
well approximated~\cite{ACRPSF} by their relativistic values computed in preceding sections. Therefore,
the contribution of the low frequency dispersive modes is given by
\be
\label{GirUV}
G_{\rm IR}(t;x,x') = 8\int_{2\pi/t}^{\om^\Lambda_{\rm U}}  |\beta_\om^{\rm Tot}|^2 d\om \times
\Phi_{\rm U}^\Lambda(x)\,  \Phi_{\rm U}^\Lambda(x') ,
\ee
which is \eq{Gir} with $\Phi_{\rm U}^m$ and $\om_{\rm U}$ replaced by $\Phi_{\rm U}^\Lambda$ and $\om^\Lambda_{\rm U}$.
The exact expression of $\beta_\om^{\rm Tot}$ in \eq{betaTot} is quite complicated.
To get an undulation, we assume $\om_L \ll \kappa \ll \om_{\rm U}^\Lambda$.
In that regime, $\beta_\om^{\rm Tot}$ is large for $\omega < T_H$, but for $\omega > T_H$ it becomes exponentially small. 
Thus, one has 
\be
\int_{2\pi/t}^{\om_{\rm U}^\Lambda}  |\beta_\om^{\rm Tot}|^2 d\om \simeq \int_{2\pi/t}^{\kappa/2\pi}  |\beta_\om^{\rm Tot}|^2 d\om.
\ee
In that range of frequencies, $\beta_\om^{\rm Tot}$ is well approximated by \eq{betatransitL}, therefore
\be
G_{\rm IR}(t;x,x') = \frac{4\kappa}\pi \left[ \text{sinh}^{-1}\left(\frac{\kappa}{2\pi \om_L} \right) - \text{sinh}^{-1}\left(\frac{2\pi}{\om_L t} \right) \right] \Phi_{\rm U}^\Lambda(x)\,  \Phi_{\rm U}^\Lambda(x').
\ee
Hence, at late times and for $\om^\Lambda_{\rm U} \gg \kappa$, 
we obtain
\be
G_{\rm IR}(t;x,x') = \frac{4\kappa}\pi \ln\left(\frac{\kappa}{\pi \om_L} \right)  \Phi_{\rm U}^\Lambda(x)\,  \Phi_{\rm U}^\Lambda(x').
\ee
When considering a BEC, the relationship between the scalar field $\phi$ and the density fluctuation
$\delta\rho$ is $\delta\rho \propto \partial_x \phi$~\cite{Balbinotetal2007}. Hence the mean value of the 
equal-time density-density two point function is
\be
\langle \partial_x \phi(x)\,  \partial_{x'} \phi(x') \rangle =
\frac{\kappa p_U^\Lambda}{\pi^2 D_L}  \ln\left(\frac{\kappa }{\pi \om_L}\right)\times
\sin \left(p_U^\Lambda x + \theta_U \right) \, \sin \left(p_U^\Lambda x' + \theta_U \right). \label{WHundul} 
\ee
This generalizes what was found in~\cite{Mayoral2011} in that, in the supersonic region, one still finds
a short distance checker board pattern in the $x, \, x'$ plane,
and the undulation amplitude still grows initially as $\ln(t)$. However, when there is a mass term,
it grows only for a finite amount of time $\sim 2\pi/\om_L$, after which it saturates.
The mass therefore provides a saturation mechanism that can occur before nonlinearities take place.
Moreover, because $p_U^\Lambda \propto \Lambda \gg \kappa$,
the r.m.s. amplitude of the undulation is large.

So far we have considered only the case where the initial state is a vacuum.
When dealing with a thermal state, as discussed in~\cite{Mayoral2011} the initial
growth rate is no longer logarithmic but linear in time. However, the mass term
acts again as an infrared regulator 
because the initial distribution of phonons
is expressed in terms of $\Om > m$, and not in terms of the constant frequency $\om$~\cite{MacherBEC}. 
Hence no divergence is found when integrating over $\om$ when computing the two point function.
In addition, the random character of the undulation amplitude is fully preserved
when taking into account some initial thermal noise.
What is modified is the  undulation r.m.s. amplitude. When the initial
temperature $T_{\rm in}$ is much larger than $T_H = \kappa/2\pi$, the above two-point function
is, roughly speaking, multiplied by $T_{\rm in}/T_H$.

\section{Conclusions}

In this paper we have studied the consequences of a mass term in the mode equation on (the analogue of) Hawking radiation.
We showed that the scattering of massive modes on a stationary black hole horizon is rather
complicated. It contains several mode mixing terms which occur on various length scales and which
interfere with each other in a non-trivial way. In what follows we summarize our main results. 

Notwithstanding the fact that the mass does not affect the near horizon mode mixing,
as can be seen in \eq{SNHR}, the mass does regularize the infrared behavior of the net $in/out$ Bogoliubov transformation
in \eq{Stotanalo}. The reason for this is the extra mode mixing,
described by $S_{\rm far}$ of \eq{Sfar}, which occurs in the supersonic inside
region, and which interferes with the near horizon scattering
so as to cancel out the divergence in $1/\om$ of the $|\beta_\om|^2$ coefficient.
Indeed, the squared norm of the
total Bogoliubov coefficient saturates as $| \beta_\om^{\rm Tot}|^2 \sim \kappa/2\pi \om_L$ for $\om \to 0$,
where $\om_L$ is the threshold frequency of \eq{2critfr}.
As a consequence, the undulation r.m.s. amplitudes
now saturate after a lapse of time $\sim 2\pi /\om_L$ and then stay constant.
This has to be contrasted with the massless case where the saturation
of the amplitude can only occur because of non-linearities, or dissipation, in the system.

The presence of a mass induces a new type of undulation in the supersonic region
that exists in black hole flows. Unlike the undulations occurring in white hole flows
which are due to some ultraviolet dispersion, this new type occurs in the hydrodynamical regime if the mass term is small enough.
It will thus appear both
in superluminal and subluminal media. However, as shown by \eq{BHundul}, the typical energy
density carried by an undulation is small, and thus this new type should be
difficult to detect.

Although the $S$-matrix coefficients governing black hole and white hole flows are the same,
in virtue of the transformation $v \to - v$ that maps one case onto the other,
the frequency ranges that contribute to the massive and the dispersive undulations are very different as can be seen by
comparing \eq{omUm} with \eq{omUL}. As a result, the white hole undulations
possess larger amplitudes. In addition, since the wave length of the undulation
is smaller in the white hole case, it gives rise to even larger amplitudes for a BEC,
 as can be seen from \eq{WHundul}. These results might also be relevant for surface waves where `transversal instabilities' have been observed~\cite{Rousseaux12}.

Finally, we emphasize that the properties of the spectrum and the undulations
depend on the mass $m$ essentially through the effective frequencies $\om_L$
and $\om_R$ of \eq{2critfr}. These frequencies are both proportional to $m$
but also depend in a non trivial way on $D_L$ and $D_R$ which determine the
spatial extension of the near horizon region, on the inside and on the outside
respectively. Therefore, as in the case of ultraviolet dispersion, these two quantities
should be conceived of as the most relevant geometrical properties, after the surface gravity $\kappa$.

\acknowledgements
We are grateful to Iacopo Carusotto for discussions and numerical simulations 
about the statistical nature of undulations in BEC.  We also thank Carlos Mayoral for collaboration at an early stage of this work, together with Xavier Busch and Yannis Bardoux for discussions.  
A.F. would like to thank the LPT Orsay for hospitality during various visits. 
This work was supported in part by the National Science Foundation under grant PHY-0856050 to Wake Forest University, as well as the ANR grant STR-COSMO, ANR-09-BLAN-0157.

\newpage

\appendix
\section{General acoustic d'Alembert equation}
\label{metricsApp}

In the hydrodynamical approximation, {\it i.e.}, when neglecting
short distance dispersion, linear density
perturbations in a moving fluid obey a $4D$ d'Alembert equation,
in a curved space-time~\cite{Unruh81,Unruh95,LivingRev,Rivista05}:
\be
\frac1{\sqrt{-g}} \partial_\mu \left( \sqrt{-g} g^{\mu \nu} \partial_\nu \phi \right) = 0.
\ee
The effective metric is
\be
ds^2 = \frac{\rho}{c} \left[ c^2 dt^2 - (dx - v_x dt)^2 - (dy - v_y dt)^2 - (dx - v_z dt)^2 \right],
\ee
where $\rho$ is the density of the fluid, $c$ the sound speed and $v$ the velocity of the background flow.

When assuming that the flow profile is one dimensional and homogeneous
in the perpendicular dimensions, the mode equation becomes effectively two dimensional
because the transverse wave number is constant. In what follows, its norm is noted
$p_{\perp}$. Assuming in addition that the flow is stationary, at fixed frequency $\om$,
the mode equation reads
\be
\left[ \frac1{\rho} (\om + i\partial_x v)\frac\rho{c^2} (\om + iv\partial_x) + \frac1{\rho} \partial_x \rho \partial_x - p_{\perp}^2 \right] \phi_\om(x) = 0 \label{pperpApp}.
\ee
In the body of the paper, we assumed $\rho(x) = c(x) = 1$.
This violates the continuity equation of fluid mechanics~\footnote{Note that this condition
does not concern surface waves because the effective metric
is different from that given above~\cite{Unruh-Schut2002,SilkePRL2010}. }
but it allows us to study exactly soluble models.

It is also interesting to see how \eq{canmodequ}
is altered by varying $c(x)$ and $\rho(x)$. Modifying the field redefinition of \eq{def_varphi}
\be
\phi_\om(x) = \frac{e^{-i \om \int^x \frac{v(x') dx'}{c^2(x')-v^2(x')}}}{\sqrt{\left|\frac{\rho(c^2-v^2)
}{c^2}\right|}} \varphi_\om(x) ,
\ee
we obtain
\be
\left[-\partial_x^2 +\left( V_G(x) + \frac{p_{\perp}^2 c^2}{c^2-v^2} - \frac{\om^2 c^2}{(c^2-v^2)^2}\right)\right] \varphi_\om(x) = 0 ,
\ee
where the effective potential is
\be
V_G(x) = \frac{\partial_x^2\sqrt{\left|\frac{\rho\, (c^2-v^2)}{c^2}\right|}}{\sqrt{\left|\frac{\rho\, (c^2-v^2)}{c^2 }\right|}} .
\ee
We see that this equation has exactly the same structure as the
simplified equation \eqref{canmodequ} that was used for some of the calculations in this paper.

\section{Hypergeometric solutions}
\label{HypApp}

We provide the exact solutions of \eq{canmodequ} for the profiles considered
in Sec.~\ref{SandroSec} and Sec.~\ref{CGHSSec}. In both cases, the solutions are
hypergeometric functions. For definiteness, we recall their definition
\be
F(a,b;c;z) = \sum_{n \in \mathbb N} \frac{\Gamma(a+n) \Gamma(b+n)\Gamma(c)}{\Gamma(a)\Gamma(b) \Gamma(c+n)} \frac{z^n}{n!},
\ee
where $a$, $b$ and $z$ are complex numbers, and $c$ is a non negative integer. Their main properties,
and the asymptotic behaviors of these functions can  be found in~\cite{AbramoSteg,Olver}.

We start with the profile of \eq{Sandrov}, and use the notation of \eq{notations}.
For $x>0$, the only ABM is
\be
\varphi_\om(x) = C (e^{\frac{2\kappa x}D}-1)^{\frac12}(1 - e^{-\frac{2\kappa x}D})^{-i\varpi} e^{-\frac{2\kappa x}D} F\left(1 - i\varpi +i \bar \Om_+, 1 - i\varpi - i\bar \Om_+ ; 2 ; e^{-\frac{2\kappa x}D} \right),\nonumber
\ee
where $C$ is an arbitrary constant.
For $x<0$, the general solution is
\begin{align}
\varphi_\om(x) =& A (1-e^{\frac{2\kappa x}D})^{\frac12}(e^{-\frac{2\kappa x}D}-1)^{i\varpi} F\left( i\varpi - i \bar \Om_+, i\varpi + i \bar \Om_+; 1 + 2i\varpi ;  1-e^{-\frac{2\kappa x}D} \right), \nonumber \\
& + B(1-e^{\frac{2\kappa x}D})^{\frac12} (e^{-\frac{2\kappa x}D}-1)^{-i\varpi} F\left(- i\varpi - i \bar \Om_+, - i\varpi + i \bar \Om_+; 1 - 2i\varpi ; 1-e^{-\frac{2\kappa x}D} \right) ,\nonumber
\end{align}
with $A$ and $B$ arbitrary constants.

For the profile in \eq{CGHSv}, and for $x>0$, the general solution is
\begin{align}
\varphi_\om(x) =& A (1-e^{-\frac{2\kappa x}D})^\frac12(e^{\frac{2\kappa x}D}-1)^{i\varpi} F\left( i\varpi - i \bar \Om, i\varpi + i \bar \Om; 1 + 2i\varpi ; 1-e^{\frac{2\kappa x}D} \right) \nonumber,\\
& + B (1-e^{-\frac{2\kappa x}D})^\frac12 (e^{\frac{2\kappa x}D}-1)^{-i\varpi} F\left(- i\varpi - i \bar \Om, - i\varpi + i \bar \Om; 1 - 2i\varpi ; 1-e^{\frac{2\kappa x}D} \right).  \nonumber
\end{align}
Here $A$ and $B$ are arbitrary constants and 
$\bar \Om = \bar \Om_>$ for $\om >\om_L$ and $ \bar \Om = i \bar \Om_<$ for $\om <\om_L$.
The definitions of these dimensionless quantities are given in Sec.~\ref{CGHSSec}.

\end{document}